\documentclass[final,3p,times]{elsarticle}
\usepackage{tabstackengine}
\usepackage{tikz}
\usepackage{tikz-3dplot}
\tdplotsetmaincoords{80}{45}
\tdplotsetrotatedcoords{-90}{180}{-90}

\tikzset{surface/.style={draw=blue!70!black, fill=blue!40!white, fill opacity=.6}}

\usepackage{epsfig,bm}
\usepackage{graphicx,amssymb}
\usepackage{amsmath}
\usepackage{physics}
\usepackage{tensor}
\usepackage{diagbox}
\usepackage{epstopdf}
\usepackage{float}
\usepackage{tabularx}
\usepackage{subfigure}
\usepackage{soul}
\usepackage{tikz-feynman}
\tikzfeynmanset{compat=1.0.0}

\usepackage[eps]{pstricks}
\usepackage[normalem]{ulem}  % 

\renewcommand\sout{\bgroup \color{red} \ULdepth=-.5ex \ULset}
\usepackage{slashed}
\newcommand{\comment}[1]{}

\begin{document}

\title{Separating the transverse and longitudinal modes of $\phi$, $\rho$, $K^*$ and $K_1$ mesons through their angular-dependent two-body decay modes}

\author[1]{In Woo Park}
\ead{darkzero37@naver.com}
\affiliation[1]{organization={Department of Physics and Institute of Physics and Applied Physics, Yonsei University},postcode={Seoul 03722},country={Korea}}

\author[2]{Hiroyuki Sako}
\ead{hiroyuki.sako@j-parc.jp}
\affiliation[2]{organization={Advanced Science Research Center, Japan Atomic Energy Agency},city={Tokai},postcode={Ibaraki 319-1195},country={Japan}}

\author[3]{Kazuya Aoki}
\ead{kazuya.aoki@kek.jp}
\affiliation[3]{orgainzation={KEK, High Energy Accelerator Research Organization,},city={Tsukuba},postcode={Ibaraki 305-0801},country={Japan}}

\author[2]{Philipp Gubler}
\ead{philipp.gubler1@gmail.com}

\author[1]{Su Houng Lee}
\ead{suhoung@yonsei.ac.kr}

\begin{abstract}
The mass shift  a spin-1 particle moving in the nuclear medium will depend on its polarization direction. To study polarization-independent mass shifts in the medium, we explore methods to isolate each polarization direction of spin-1 mesons through the angular-dependent two-body decay modes.  Specifically, we study   $\phi\to K^{+}K^{-}$, $\rho\to\pi\pi$, $K^{*}\to K\pi$, $\phi\to e^{+}e^{-}$ and $K_{1}\to\rho K(K^{*}\pi)$ decays. 
While each polarization mode can be isolated through angular dependencies, accomplishing the decomposition for the $K_1$  meson further requires measuring the polarization of the $\rho(K^*)$ meson.  Concerning $K^{*}$ and $K_1$ mesons, since both particles have vacuum widths smaller than 100 MeV, they are ideal candidates for experimentally measuring chiral partners. The simultaneous observation of mass shifts of these chiral partners would provide valuable insights into the contribution of chiral symmetry breaking to the generation of hadron masses.
\end{abstract}

\maketitle

\section{Introduction}\label{sec:introduction}
A fundamental problem in quantum chromodynamics (QCD) is understanding the generation of hadron masses, which are believed to 
be closely related to the spontaneous chiral symmetry breaking in vacuum \cite{Nambu:1961tp,Nambu:1961fr,Hatsuda:1985eb,Brown:1991kk,Hatsuda:1991ez,Leupold:2009kz,Gubler:2014pta}. 
In a heavy ion collision, when the collision energy is large enough, the initial temperature is expected to be well above the region where chiral symmetry is restored. Hence, careful study of effects that reflect the initial stages is expected to reveal effects of chiral symmetry restoration.  Also nuclear target experiments can reveal hadron properties from inside nuclear matter where chiral symmetry is expected to be partially restored. 
To probe the mass shift in these environments, several experiments have been performed for more than a decade \cite{Hayano:2008vn,JPARC:2023quf,Metag:2017yuh,Ohnishi:2019cif,Salabura:2020tou}. 
On general grounds, it is well established that the longitudinal and transverse modes of vector fields can exhibit different behaviors when 
propagating through a hot and/or dense nuclear medium (see, for instance, Ref.~\cite{LeBellac:1996} and the references cited therein for discussions within thermal field theory). 
In fact, the expected mass shifts for the longitudinal and transverse modes of the vector meson go in opposite directions 
so that accummlating all data will induce a spread in the invariant mass distribution not originating in the increase in the width \cite{Lee:1997zta,Kim:2019ybi}. 
Since the chiral symmetry breaking effect is largely a momentum-independent component of the mass shift, it is important to study mesons emanating from dense matter at slow velocities \cite{Lee:2023ofg}.

The J-PARC E16 experiment \cite{JPARC:2023quf, Aoki:2023qgl} is performing experiments to observe the mass shift of the $\phi$ meson through $e^+e^-$ pairs produced in pA collisions. The previous KEK experiment  \cite{KEK-PS-E325:2005wbm} revealed that when observing $\phi$ mesons with small velocity ($\beta\gamma<1.25$), a mass shift was observed. The present E16 experiment will have sufficient statistics to confirm the mass shift for slowly moving $\phi$ mesons. The J-PARC E88 experiment \cite{Sako} will supplement E16 by measuring much higher-statistics $\phi$ mesons through the $K^+K^-$ decay, which enables the analysis of the $\phi$ meson mass dependence on momentum and polarization. Because the $\phi$ meson has a small vacuum width, the expected increase of the width in the nuclear medium will be small enough
to allow for a successful reconstruction of the peak position.

One can concentrate on the effects due to the chiral symmetry restoration in a medium, if one studies the mass shifts of chiral partners, as the mass shifts between these particles will only depend on chiral symmetry breaking. 
Since both chiral partners have to be observed, focusing on chiral partners with small widths is crucial. 
This naturally leads one to study the $K^*,K_1$ system \cite{Lee:2019tvt,Song:2018plu}. 
The purpose of this paper is to illuminate how the two transverse and one longitudinal polarizations of the vector (axial vector) meson can be disentangled from the angular distributions of its two-body decays. The method can be applied to ongoing nuclear target experiments to isolate the mass shift for both the transverse and longitudinal components, thereby identifying the momentum and polarization independent component of the mass shift. 
The method used here can also be applied to other massive spin-1 particles with minimum modifications. 
For example, hadronic polarization phenomena have recently drawn attention. Specifically, the $\Lambda$ baryon polarization in 
noncentral heavy-ion collisions was observed  \cite{STAR:2017ckg,STAR:2018gyt}. 
The polarization of $\Lambda$ baryons has demonstrated the presence of strong vorticity in such collisions, 
typically defined relative to the direction perpendicular to the event plane. Similar measurements conducted at the same and other facilities \cite{STAR:2021beb,HADES:2021Kornas} have been extended to different baryon species and, more recently, to vector particles \cite{STAR:2020xbm}. Although many theoretical studies are currently exploring the mechanisms behind the varying degrees of polarization across different channels, our  formalism could provide extra ingredients to the puzzle by identifying the polarization effects through the longitudinal and transverse components defined relative to the propagating vector(axial vector)-meson direction.

Our findings show that for the $\phi$ decay in the forward direction, the angular dependence is dominated by the transverse component in the $\phi\to e^{+}e^{-}$ channel, whereas the longitudinal mode dominates in the $\phi\to K^{+}K^{-}$, $\rho\to\pi\pi$ and $K^*\to K\pi$ channels \cite{Park:2022ayr}. 
This shows that it is important to perform complementary measurements using both the dilepton and pseudoscalar meson decay channels. For the $K_1\to\rho K(K^{*}\pi)$ decay, the separation of the two modes can be achieved by further examining the polarization of the vector mesons in the decay modes \cite{Park:2024vga}. All these features can be understood from simple arguments using conserved angular momentum and rotations, making use of Wigner $D$ matrices as discussed in \cite{Faccioli:2010kd} and \cite{Schilling:1969um}.

The paper is organized as follows. In Sec.\ref{sec:config}, we introduce the two transverse and one longitudinal polarization of a massive spin-1 particle along with the spin density matrix. In Sec.\ref{sec:decayrate}, we present the partial decay widths of each decay channel and the corresponding effective interaction Lagrangians, which are used to evaluate the coupling constants for each decay. Next, we introduce the three helicity states of a spin-1 particle and analyze how the general angular distribution can be obtained from each interaction Lagrangian. In Sec.\ref{sec:wigmatrix}, we reproduce the same result obtained in the previous section using the helicity formalism. Finally, we summarize our discussion and conclude the paper in Sec.\ref{sec:sumandconc}. Detailed calculations are given in the appendices.

\section{General configuration of the initial spin-1 meson and its spin density matrix}\label{sec:config}

In general, the initial spin-1 state can be written as 
\begin{align}
   \ket{V} = \displaystyle\sum_{\lambda=\pm1,0}a_{\lambda}\ket{\lambda}.
   \label{eq:vec_meson_config}
\end{align}
Here, $\ket{\lambda}$ stands for the three different polarizations of the initial vector (axial vector) meson state, 
($\lambda = \pm 1$: transverse polarization, $\lambda = 0$: longitudinal polarization), and 
 $a_{\lambda}$ is the amplitude defining the strength of each mode. 
Further details concerning 
the four-polarization vector for each $\lambda$ are given in Appendix \ref{appendix:basis}. 
Using Eq.(\ref{eq:vec_meson_config}), we can define the spin density matrix  $\rho_{\lambda\lambda^{\prime}}=a_{\lambda}a_{\lambda^{\prime}}^{\star}$ of the initial spin-1 meson. Its trace 
is normalized as $\rho_{11}+\rho_{00}+\rho_{-1-1}=1$. A transversely polarized spin-1 state in which the $z$ component of the total angular momentum is $J_{z}=\pm1$, will have $\rho_{00}=0$. Conversely, a longitudinal polarization (with $J_{z}=0$) corresponds to $\rho_{00}=1$. 
For an unpolarized spin-1 state, all the diagonal entries of the density matrix will have $\rho_{11}=\rho_{00}=\rho_{-1-1}=\frac{1}{3}$.

\section{$\phi,\;\rho,\;K^{*},\;K_{1}$ meson decay rate and their angular dependence}\label{sec:decayrate}
In this section, we discuss the decay of the vector mesons $\phi$, $\rho$, $K^*$ and the axial vector meson $K_1$, which can be categorized into 3 different cases. The first is the vector meson ($V$) decaying into two pseudoscalar-mesons ($P$). Second is the vector meson ($V$) decaying into a positron and an electron. Third is the axial vector meson ($A$) decaying into a vector meson ($V$) and a pseudoscalar meson ($P$).

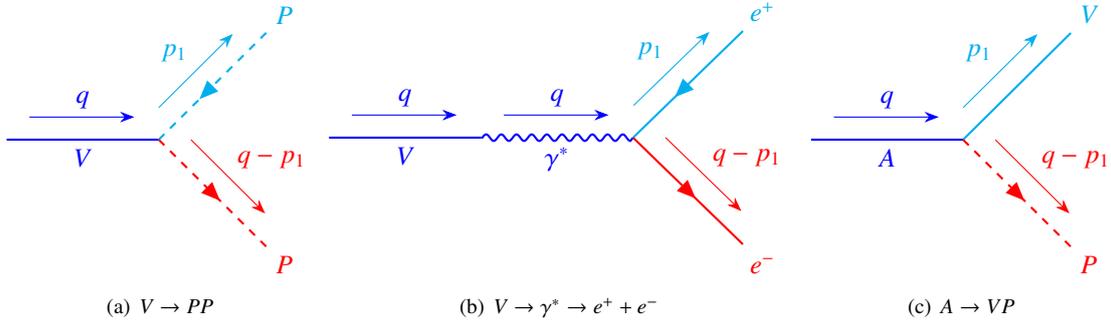
\begin{figure}[H]
\centering

\subfigure [$V\to PP$]{
\begin{tikzpicture}
\begin{feynman}[large]
\vertex (a) ;
\vertex [right=of a] (b);
\vertex [above right=of b] (f1){\textcolor{cyan}{\(P\)}};
\vertex [below right=of b] (f2){\textcolor{red}{\(P\)}};
\diagram* {
(a) -- [blue,edge label'=\textcolor{blue}{\(V\)},momentum={[arrow style=blue]\(q\)}] (b) , (b) -- [cyan,anti charged scalar,momentum={[arrow style=cyan]\(p_{1}\)}] (f1), (b) -- [red,charged scalar,momentum={[arrow style=red]\(q-p_{1}\)}] (f2)};
\end{feynman}
\end{tikzpicture}
\label{fig:phitokaon}
}\subfigure [$V\to\gamma^{*}\to e^{+}+e^{-}$]{
\begin{tikzpicture}
\begin{feynman}[large]
\vertex (a) ;
\vertex [right=of a] (b);
\vertex [right=of b] (c);
\vertex [above right=of c] (f1){\textcolor{cyan}{\(e^{+}\)}};
\vertex [below right=of c] (f2){\textcolor{red}{\(e^{-}\)}};
\diagram* {
(a) -- [blue,edge label'=\textcolor{blue}{\(V\)},momentum={[arrow style=blue]\(q\)}] (b) , (b) -- [blue,photon,edge label'=\textcolor{blue}{\(\gamma^{*}\)},momentum={[arrow style=blue]\(q\)}] (c), (c) -- [cyan,anti fermion,momentum={[arrow style=cyan]\(p_{1}\)}] (f1), (c) -- [red,fermion,momentum={[arrow style=red]\(q-p_{1}\)}] (f2)};
\end{feynman}
\end{tikzpicture}
\label{fig:phitodilepton}}
\subfigure [$A\to VP$]{
\begin{tikzpicture}
\begin{feynman}[large]
\vertex (a) ;
\vertex [right=of a] (b);
\vertex [above right=of b] (f1){\textcolor{cyan}{\(V\)}};
\vertex [below right=of b] (f2){\textcolor{red}{\(P\)}};
\diagram* {
(a) -- [blue,edge label'=\textcolor{blue}{\(A\)},momentum={[arrow style=blue]\(q\)}] (b) , (b) -- [cyan,momentum={[arrow style=cyan]\(p_{1}\)}] (f1), (b) -- [red,charged scalar,momentum={[arrow style=red]\(q-p_{1}\)}] (f2)};
\end{feynman}
\end{tikzpicture}\label{fig:AtoVP}}

\caption{Tree-level diagram of each decay process (a) $V\to PP$, (b) $V\to\gamma^{*}\to e^{+}e^{-}$ and (c) $A\to VP$.}
\label{fig:feyndiagram}
\end{figure}

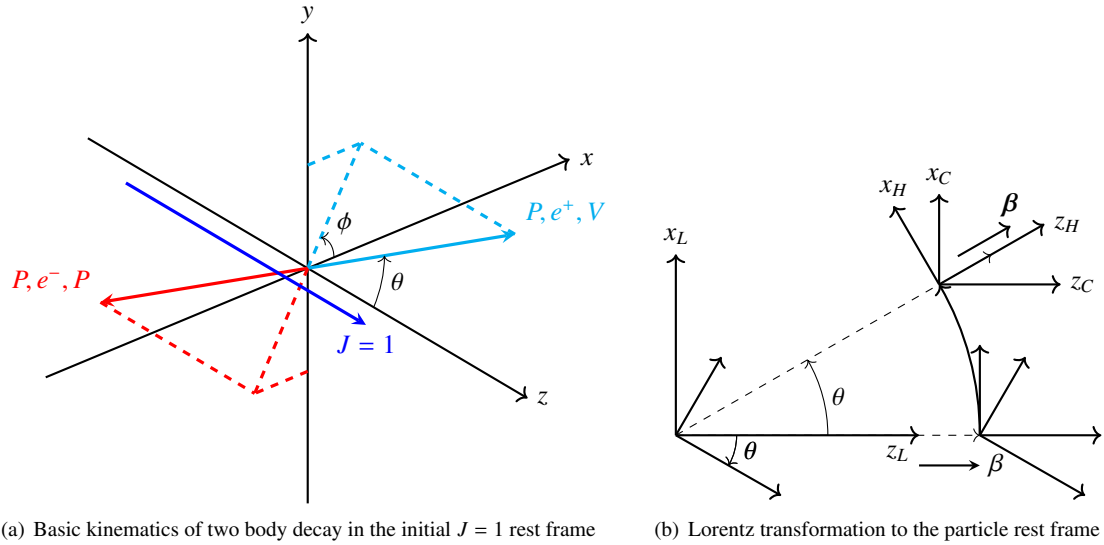
\begin{figure}[H]
\subfigure[Basic kinematics of two body decay in the initial $J=1$ rest frame]{
\tdplotsetmaincoords{60}{110}
\begin{tikzpicture}
[scale=4.5,tdplot_main_coords,rotate around z=0,rotate around x=-90,rotate around z=180,rotate around y=-30]
 % variables
  \def\rvec{1}
  \def\thetavec{30}
  \def\phivec{60}
 
  % axes
  \coordinate (O) at (0,0,0);
  \draw[thick,->] (-1.0,0,0) -- (1.0,0,0) node[right]{$x$};
  \draw[thick,->] (0,-0.8,0) -- (0,0.8,0) node[above]{$y$};
  \draw[thick,->] (0,0,-1.0) -- (0,0,1.0) node[right]{$z$};

\tdplotsetcoord{P}{0.8\rvec}{\thetavec}{\phivec}
  \draw[-stealth,very thick,cyan] (O)  -- (P) node[above right] {$P,e^{+},V$};
  \draw[dashed,very thick,cyan]   (O)  -- (Pxy);
  \draw[dashed,very thick,cyan]   (P)  -- (Pxy);
  \draw[dashed,very thick,cyan]   (Py) -- (Pxy);
  
\tdplotsetcoord{Q}{0.8\rvec}{-\thetavec+180}{\phivec+180}
  \draw[-stealth,very thick,red] (O)  -- (Q) node[above left] {$P,e^{-},P$};
  \draw[dashed,very thick,red]   (O)  -- (Qxy);
  \draw[dashed,very thick,red]   (Q)  -- (Qxy);
  \draw[dashed,very thick,red]   (Qy) -- (Qxy);
%\tdplotdrawarc[coordinate system, draw styles]{center}{r}{angle start}{angle end}{label options}{label}

  \tdplotdrawarc[->]{(0,0,0)}{0.1}{0}{\phivec}{above=2pt,right=-1pt,anchor=south west}{$\phi$}
  
  \tdplotdrawarc[->,rotate around z=\phivec-90,rotate around y=-90]{(0,0,0)}{0.3}{0}{\thetavec}
    {anchor=west}{$\theta$}

\tdplotsetcoord{R}{0.6\rvec}{180-26.5}{\phivec+90}

\tdplotsetcoord{S}{0.6\rvec}{26.5}{\phivec+90}
\draw[-stealth,very thick,blue] (R)  -- (S) node[below] {$J=1$};
\end{tikzpicture}
\label{fig:kinematics}
}
\subfigure[Lorentz transformation to the particle rest frame]{
\tdplotsetmaincoords{0}{0}
\begin{tikzpicture}
[scale=4]
 % variables
\def\R{0.6}
  % axes
  \coordinate (O) at (0,0,0);
  \draw[thick,->] (0,0,0) -- (0.8,0,0) node[below left]{$z_L$};
  \draw[thick,->] (0,0,0) -- (0,0.6,0) node[above]{$x_L$};
  \draw[->,black,dashed](0,0,0) -- (1,0,0);
    \draw[->,black,dashed](0,0,0) -- ({1.2*cos(30)},{1.2*sin(30)},0)node[anchor=180+10,black]{};
%\tdplotdrawarc[coordinate system, draw styles]{center}{r}{angle start}{angle end}{label options}{label}

\tdplotdrawarc[->]{(0,0,0)}{0.5}{0}{30}{anchor=west}{$\theta$}
\tdplotdrawarc[->,thick]{(0,0,0)}{1}{0}{30}{anchor=west}{}

\coordinate (R) at (0.8,-0.1,0);
\coordinate (S) at (1,-0.1,0);

\draw[-stealth,thick,black] (R)  -- (S) node[right] {$\beta$};

\coordinate (P) at (1,0,0);
\coordinate (Q) at ({cos(30)},{sin(30)},0);

\tdplotsetrotatedcoords{0}{0}{-30}
\tdplotsetrotatedcoordsorigin{(O)}
\draw[thick,tdplot_rotated_coords,->] (0,0,0) -- (0.4,0,0)node[right]{};
\draw[thick,tdplot_rotated_coords,->] (0,0,0) -- (0,0.3,0)node[above]{};
\tdplotdrawarc[->]{(0,0,0)}{0.2}{0}{-30}{anchor=west}{$\theta$}

\tdplotsetrotatedcoords{0}{0}{-30}
\tdplotsetrotatedcoordsorigin{(P)}
\draw[thick,tdplot_rotated_coords,->] (0,0,0) -- (0.4,0,0)node[right]{};
\draw[thick,tdplot_rotated_coords,->] (0,0,0) -- (0,0.3,0)node[above]{};
\tdplotdrawarc[->]{(0,0,0)}{0.2}{0}{-30}{anchor=west}{$\theta$}

\tdplotsetrotatedcoords{0}{0}{0}

\tdplotsetrotatedcoordsorigin{(P)}
\draw[thick,tdplot_rotated_coords,->] (0,0,0) -- (0.4,0,0)node[right]{};
\draw[thick,tdplot_rotated_coords,->] (0,0,0) -- (0,0.3,0)node[above]{};

\tdplotsetrotatedcoords{0}{0}{30}
\tdplotsetrotatedcoordsorigin{(Q)}
\draw[thick,tdplot_rotated_coords,->] (0,0,0) -- (0.4,0,0)node[right]{$z_H$};
\draw[thick,tdplot_rotated_coords,->] (0,0,0) -- (0,0.3,0)node[above]{$x_H$};
\draw[thick,tdplot_rotated_coords,->](0.1,0.05,0) -- (0.3,0.05,0)node[above]{$\bm{\beta}$};

\tdplotsetrotatedcoords{0}{0}{0}
\tdplotsetrotatedcoordsorigin{(Q)}
    \draw[thick,tdplot_rotated_coords,->] (0,0,0) -- (0.4,0,0)node[right]{$z_C$};
\draw[thick,tdplot_rotated_coords,->] (0,0,0) -- (0,0.3,0)node[above]{$x_C$};

\end{tikzpicture}\label{fig:helframe}
}
\caption{Sketch of (a) Basic kinematics of two body decay in the initial $J=1$ rest frame, (b) Lorentz transformation of a system from the Lab frame to the helicity rest frame and the canonical rest frame of a particle when azimuthal angle is set to $\phi=0$.}
\end{figure}

We adopt phenomenological interaction Lagrangians for each decay process to calculate the respecitive amplitudes.
The tree diagrams for the considered decay processes are given in  Fig.\ref{fig:feyndiagram}. 
Fig.\ref{fig:kinematics} displays the kinematics of the two body decay of a spin-1 meson in its rest frame with the polar and azimuthal angles $\theta$ and $\phi$ of the produced particle (cyan line). The blue line in Fig.\ref{fig:kinematics} depicts the momentum of the initial spin-1 meson in the Lab frame. The helicity rest frame of a produced particle is established through a Lorentz boost along $z$-direction, 
followed by an Euler rotation $R(\phi,\theta,0)$, where $R(\alpha,\beta,\gamma)=e^{-i\alpha J_{3}}e^{-i\beta J_{2}}e^{-i\gamma J_{3}}$. $R(\alpha,\beta,\gamma)$ is a rotation about the fixed $z$-axis by an angle of $\gamma$, which is followed by a rotation about the fixed $y$-axis by an angle of $\beta$, and finally followed by a rotation about the fixed $z$-axis by an angle of $\alpha$. 
On the other hand, the canonical rest frame of a particle is reached by an inverse Euler rotation $R^{-1}(\phi,\theta,0)$ followed by a Lorentz boost along the $z$-axis, finally followed by an Euler rotation $R(\phi,\theta,0)$. 
Fig.\ref{fig:helframe} illustrates the Lorentz transformation of a system from the Lab frame to the helicity rest frame and the canonical rest frame of a particle when the azimuthal angle of a produced particle is fixed to $\phi=0$. 
The subscripts $L$, $H$ and $C$ stand for Lab frame, Helicity rest frame and Canonical rest frame, respectively.

As we will see in the following subsections, the polar angular distributions of $V\to PP$, $V\to ll$ and $A\to VP$ can be expressed using the universal formula 
\begin{align}\label{eq:polarangdist}
\dv{\Gamma}{\cos\theta}&=\frac{1}{16\pi}\frac{\abs{\bm{p_1}}}{E_{\text{c.m.}}^{2}}\left((1-\rho_{00})\abs{\mathcal{M}}_{T}^{2}+\rho_{00}\abs{\mathcal{M}}_{L}^{2}\right),
\end{align}
where $\abs{\mathcal{M}}_{T}^{2}$ and $\abs{\mathcal{M}}_{L}^{ 2}$ are the transverse and longitudinal components of the corresponding decay amplitude, respectively. $\abs{\bm{p_{1}}}=\frac{1}{2m_{0}}\sqrt{\left(m_{0}^{2}-(m_{1}-m_{2})^{2}\right)\left(m_{0}^{2}-(m_{1}+m_{2})^{2}\right)}$\label{eqn:cmmomentum} is the momentum of both produced particles in the c.m. frame and $E_{\text{c.m.}}$ is the total energy. The above formula clarifies the role of each polarization mode in the angular decay distribution. The angular dependence of   $\dv{\Gamma}{\cos\theta}$ for a transversely or longitudinally polarized mode can be calculated by substituting $\rho_{00}=0$ or $\rho_{00}=1$ in Eq.(\ref{eq:polarangdist}), respectively. More details on $\abs{\mathcal{M}}_{T}^{2}$ and $\abs{\mathcal{M}}_{L}^{2}$ for each processes are given in Appendix.\ref{appendix:angular_dep_comp}.

\subsection{$V\to PP$}\label{sec:K+K-}

For the $V\to PP$ decay, we are going to study the following decays: $\phi\to K^{+}K^{-}$, $\rho\to\pi\pi$ and $K^{*}\to K\pi$. Partial decay widths are taken from the Particle Data Group \cite{Workman:2022ynf} and are given as
\begin{align}\label{eq:vppdecaywidth}
\Gamma_{\phi K^+K^-}&=\frac{g_{\phi K^+K^-}^{2}}{8\pi}\frac{\abs{\bm{p_{1}}}}{m_{\phi}^{2}}|\frac{4}{3}\abs{\bm{p_1}}^{2}=4.249\times(0.492)\;\text{MeV},\nonumber\\
\Gamma_{\rho\pi\pi}&=\frac{g_{\rho\pi\pi}^{2}}{8\pi}\frac{\abs{\bm{p_{1}}}}{m_{\rho}^{2}}\frac{4}{3}\abs{\bm{p_{1}}}^{2}=149\,\text{MeV},\\
\Gamma_{K^{*}K\pi}&=\frac{g_{K^{*}K\pi}^{2}}{8\pi}\frac{\abs{\bm{p_{1}}}}{m_{K^{*}}^{2}}4\abs{\bm{p_{1}}}^{2}=51.4\,\text{MeV}.\nonumber
\end{align} 
$m_{0}$ is the mass of the initial particle and $m_{1}$ and $m_{2}$ are that of the final particles. 
The number inside the bracket is the branching ratio of $\phi\to K^{+}K^{-}$. The coupling constants $g_{\phi K^+K^-}$, $g_{\rho\pi\pi}$ and $g_{K^{*}K\pi}$ and $\abs{\bm{p_{1}}}$ corresponding to each decay process are given in Table \ref{table:moment_coupl}. 
For the mass values, we use 
$m_{\phi}=1019.461\;\text{MeV}$ and $m_{K^{\pm}}=493.677\;\text{MeV}$ in $\phi\to K^{+}+K^{-}$. 
For the $\rho\to\pi+\pi$ and $K^*\to K+\pi$ decay, $K^*$ and $K$ masses are isospin averaged using the PDG data \cite{Workman:2022ynf}, giving $m_{K}=495.644$ MeV, $m_{K^{*}}=893.61$ 
MeV, $m_{\rho}=775.16$ MeV and $m_{\pi}=138.037$ MeV. 
To evaluate the decay widths, we take the average of the initial 3 spin degrees of freedom and sum over the final spin. $\rho\to\pi\pi$ has 3 isospin dependent decay modes, which are $\rho^{+}\to\pi^{+}\pi^{0},\,\rho^{-}\to\pi^{-}\pi^{0}$ and $\rho^{0}\to\pi^{+}\pi^{-}$. 
Similarly, $K^{*}\to K\pi$ has 4 isospin dependent decay modes, which are $K^{+*}\to K^{+}\pi^{0}$, $K^{0}\pi^{+}$ and $K^{0*}\to K^{0}\pi^{0}$, $K^{+}\pi^{-}$. 
Therefore, the initial isospin average is obtained by dividing by a factor of 3 for $\rho\to\pi\pi$ and 4 for $K^{*}\to K\pi$. 

The interaction Lagrangians of the vector-meson with two pseudoscalar-mesons are taken from  \cite{Klingl:1996by,Kaymakcalan:1984bz,Meissner:1987ge,Sung:2021myr} and given as 
\begin{align}\label{eq:vpplagrangian}
&\mathcal{L}_{\phi K^+K^-}=g_{\phi K^+K^-}(K^{+}\partial_{\mu}K^{-}-K^{-}\partial_{\mu}K^{+})\phi^{\mu},\nonumber\\
&\mathcal{L}_{\rho\pi\pi}=g_{\rho\pi\pi}\bigg(\pi^{+}\overset{\leftrightarrow}{\partial}_\mu\pi^{-}\rho_{0}^{\mu}+\pi^{+}\overset{\leftrightarrow}{\partial}_\mu\pi^{0}\rho_{-}^{\mu}+\pi^{-}\overset{\leftrightarrow}{\partial}_\mu\pi^{0}\rho_{+}^{\mu}\bigg),\\
&\mathcal{L}_{K^{*}K\pi}=\sqrt{2}g_{K^{*}K\pi}\bigg(\bar{K}\vec{\tau}\vdot\partial_{\mu}\vec{\pi}-\partial_{\mu}\bar{K}\vec{\tau}\vdot\vec{\pi}\bigg)K^{*\mu}.\nonumber
\end{align}
Here, $\phi^{\mu}$, $\rho^{\mu}$ and $K^{*\mu}$ are the vector-meson fields corresponding to $\phi$, $\rho$ and $K^*$ mesons, respectively. $K^{+},\;K^{-}$ are charged kaon fields and $K^{*},\,K$ are isodoublet fields whose matrix representations are listed in Appendix \ref{appendix:effectivelagrangian}. 
We further define $\pi^{+}\overset{\leftrightarrow}{\partial_\mu}\pi^{-}=\pi^{+}(\partial_{\mu}\pi^{-})-(\partial_{\mu}\pi^{+})\pi^{-}$. 

\begin{figure}
\centering
\subfigure [$\dv{\Gamma}{\cos\theta}$ of $\phi\to K^{+}K^{-}$ decay in the c.m. frame]{
\includegraphics[width=3.1in,height=1.6in]{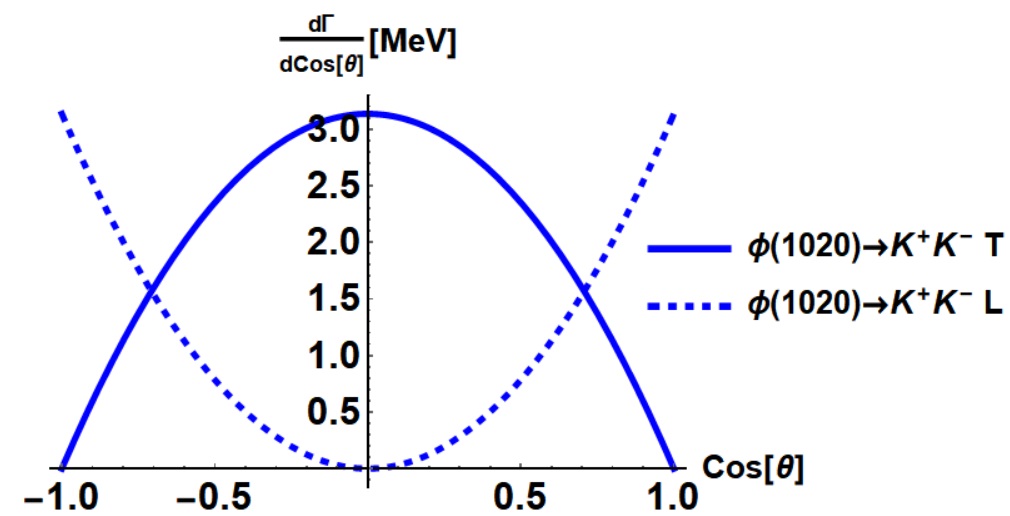}
\label{fig:kaon_cm_costheta_distribution}
}\\
\subfigure [$\dv{\Gamma}{\cos\theta}$ of $\rho\to\pi\pi$ in the c.m. frame]{
\includegraphics[width=3.1in,height=1.6in]{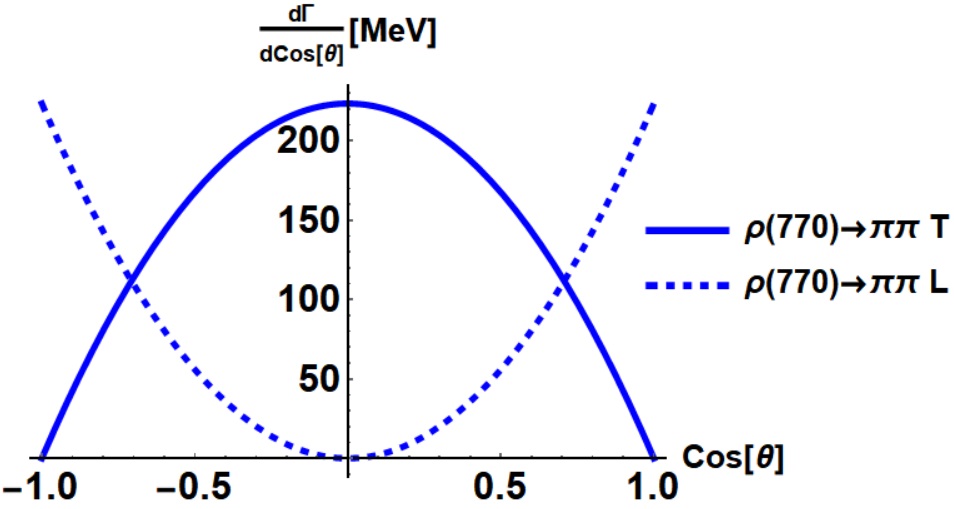}
\label{fig:rhodecay}
}\subfigure [$\dv{\Gamma}{\cos\theta}$ of $K^{*}\to K\pi$ in the c.m. frame]{
\includegraphics[width=3.1in,height=1.6in]{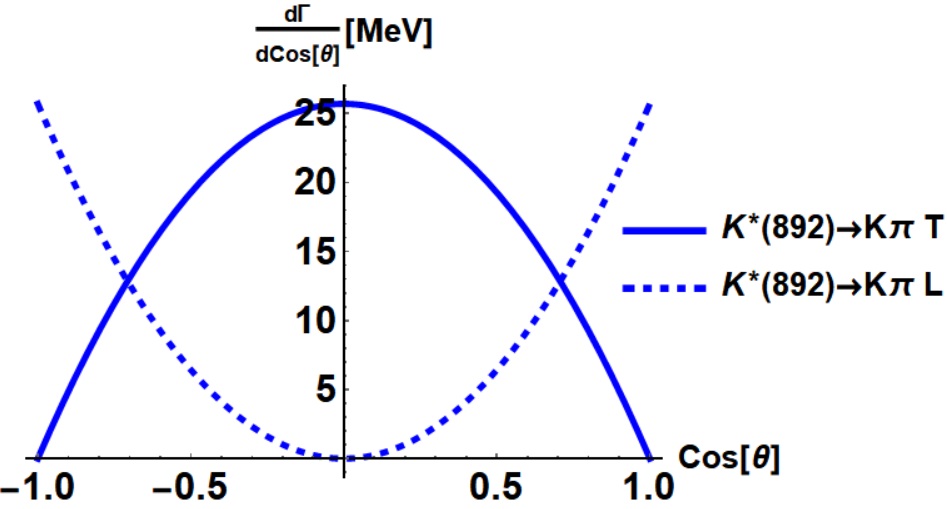}
\label{fig:kstardecay}
}
\caption{Angular distribution of decay rate of (a) $\phi\to K^{+}+K^{-}$ and (b) $\phi\to e^{+}+e^{-}$ (c) $\rho\to\pi\pi$ and (d) $K^{*}\to K\pi$ in the c.m. frame for each polarization. T stands for transverse polarization and L stands for longitudinal polarization.}
\label{fig:cos_theta_distribution}
\end{figure}

We now study the angular dependence of the decay for each mode, which can be obtained by multiplying the 
projection operators as given in the Appendix as Eq.(\ref{eq:projection_op}) 
to the tensor-valued decay amplitudes of Eq.(\ref{eq:decaytensor}) and finally contracting the Lorentz indices. 
This procedure leads to the results given in Eq.(\ref{eq:tlamplitude}). 
The corresponding angular distributions for each polarization mode of $\phi\to K^{+}K^{-}$, $\rho\to\pi\pi$ and $K^{*}\to K\pi$ are shown in Figs.\ref{fig:kaon_cm_costheta_distribution}, \ref{fig:rhodecay}, \ref{fig:kstardecay}, respectively.

If the initial vector-meson is in the general configuration given in Eq.(\ref{eq:vppamp}), the general normalized angular distribution for  
all $V\to PP$ decays can be expressed as
\begin{align}\label{eq:vppangledist}
\frac{1}{\Gamma}\dv{\Gamma}{\Omega}&=\frac{3}{8\pi}\bigg(2\rho_{00}\cos^{2}\theta+(1-\rho_{00})\sin^{2}\theta-2\text{Re}[\rho_{1-1}]\sin^{2}\theta\cos2\phi+2\text{Im}[\rho_{1-1}]\sin^{2}\theta\sin2\phi\\
&-\sqrt{2}\text{Re}[\rho_{10}-\rho_{-10}]\sin2\theta\cos\phi+\sqrt{2}\text{Im}[\rho_{10}+\rho_{-10}]\sin2\theta\sin\phi\bigg).\nonumber
\end{align}
Here, $\theta$ and $\phi$ are explained at the beginning of this section. As we can see, the normalized angular distribution of $V\to PP$ decay no longer has any mass or coupling constant dependence. More details about the calculation of the general angular distribution is given in 
Appendix \ref{appendix:kaon_distribution}. Integrating $\frac{1}{\Gamma}\dv{\Gamma}{\Omega}$ over $\phi$ from 0 to $2\pi$ yields the 
$\theta$ distribution 
\begin{align}\label{eq:Wtheta1}
\frac{1}{\Gamma}\dv{\Gamma}{\cos\theta}&=\frac{3}{4}\left(1-\rho_{00}+(3\rho_{00}-1)\cos^{2}\theta\right).
\end{align}
Here, multiplying the above equation by $\Gamma$, and using the matrix element given in the Appendix.\ref{appendix:polarization_tensor}, this equation can be cast into the general form given in Eq.(\ref{eq:polarangdist}). 
The coupling constants $g_{\phi K^+K^-},\;g_{\rho\pi\pi}$ and $g_{K^{*}K\pi}$ are determined by the strength of the hadronic decay. Higher order terms of the chiral expansion will yield different forms of vertices, which commonly have more derivatives. 
Nonetheless, regardless of the structure of the vertices, we find that the decay amplitude can always be expressed as a term proportional to $F(m_{0},m_{1},m_{2}) \times \epsilon_{\lambda}\cdot (p_1 - p_2)$. This is so because of 
$p_2 = q - p_1$, $\epsilon_{\lambda}\cdot q = 0$, and the initial particle and the two decaying particles are on mass shell, ie. $q^2=m_0^2$, $p_1^2=m_1^2$ and $p_2^2=m_2^2$, respectively. 
This leads to the functional form of $F(m_{0},m_{1},m_{2})$, which is not dictated by the specific structure of the vertex, 
but solely depends on the masses of the initial and final particles. 
Therefore it can be included into the definition of $g_{VPP}$ without changing the angular distribution.

\begin{table}[t]
\caption{Coupling constants and momenta of the produced particle in the c.m. frame for each decay process.}
\label{table:moment_coupl}
\centering
 \begin{tabular}{c | c | c | c | c | c | c }
    \hline
    Decay & $\phi\to K^{+}K^{-}$ & $\phi\to e^{+}e^{-}$ & $\rho\to\pi\pi$  & $K^{*}\to K\pi$ &  $K_{1}\to\rho K$ & $K_{1}\to K^{*}\pi$
    \\
    \hline
    $\abs{\bm{p_{1}}}$(MeV)& $127$& $510$ & $362$   & $289$ &   $27$ & $299$    \\
   \hline
  $g_{ABC}$ & $4.48$& $13.4$ &$5.96$  &  $3.27$  &  $3.26$  &  $0.71$  \\
  \hline
\end{tabular}
\end{table}

We can see in Fig.\ref{fig:kaon_cm_costheta_distribution} that the transverse amplitude vanishes at  $\theta=0$ or $\pi$. This can be inferred from the structure of the interaction Lagrangian in Eq.(\ref{eq:vpplagrangian}), from which it is understood that the hadronic decay amplitude 
of the vector meson in its rest frame is proportional to the dot product of the relative momentum of the final pseudoscalar mesons and the polarization vector of the initial vector meson, $(\bm{p}_{1}-\bm{p}_{2}) \vdot\bm{\epsilon}$. 
The relative momentum and transverse polarization vectors are orthogonal to each other at $\theta=0,\pi$, as the relative momentum is along the $z$-axis whereas the transverse polarization vectors of the initial vector meson lie in the $xy$ plane. Detailed calculations regarding the disentanglement of polarization using the polarization tensor of the transverse and longitudinal modes are studied in Appendix \ref{appendix:polarization_tensor}.

\subsection{$V\to e^{+}+e^{-}$}

The partial decay width of the $\phi\to e^{+}e^{-}$ decay, taken from the PDG \cite{Workman:2022ynf} is given as
\begin{align}\label{eq:vlldecaywidth}
\Gamma_{\phi e^+e^-}&=\frac{1}{8\pi}\frac{\abs{\bm{p_1}}}{m_{\phi}^{2}}\frac{64\pi^{2}\alpha^{2}}{3g_{J}^{2}}m_{\phi}^{2}=4.249\times(2.974\times10^{-4})\;\text{MeV},
\end{align}
where the number inside the bracket is the branching ratio and $\alpha=e^{2}/4\pi$ is the fine structure constant. 
$g_{J}$ is given as the coupling constant between the $\phi$ meson and the photon. $\abs{\bm{p_{1}}}$ and $g_{J}$ are listed in Table \ref{table:moment_coupl}.

\begin{figure}
\centering
\includegraphics[width=3.2in,height=1.7in]{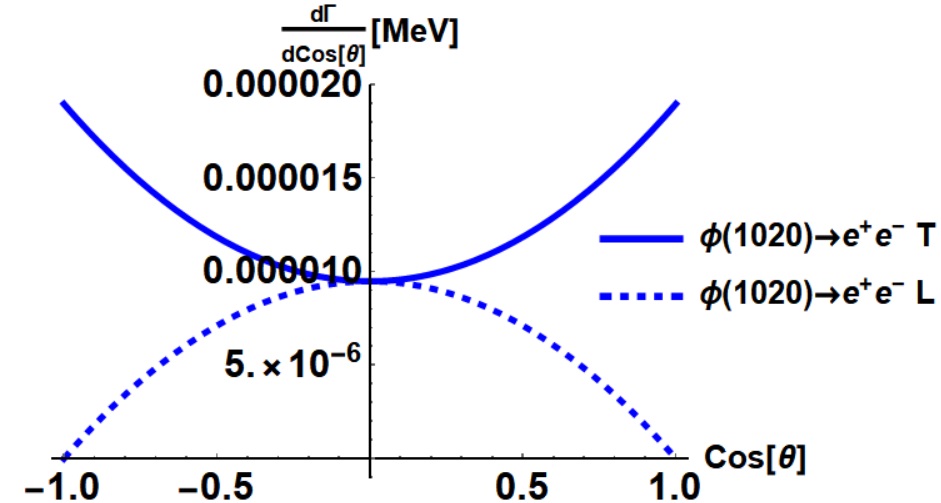}
\caption{$\dv{\Gamma}{\cos\theta}$ of $\phi\to e^{+}e^{-}$ in the c.m. frame}\label{fig:dilepton_cm_costheta_distribution}
\end{figure}

In order to describe the $\phi$ meson decaying into a pair of leptons, we employ the vector meson dominance (VMD) model and use the interaction Lagrangian described in \cite{Klingl:1996by,sakurai1969currents}, which is given as
\begin{align} \label{eq:vlllagrangian}
&\mathcal{L}_{\phi\gamma}=-\frac{e}{2g_{J}}F^{\mu\nu}\phi_{\mu\nu},\quad\mathcal{L}_{\gamma e^+e^-}=-e\bar{\psi}\gamma^{\mu}A_{\mu}\psi.
\end{align}
Field tensors are defined as $F^{\mu\nu}=\partial^{\mu}A^{\nu}-\partial^{\nu}A^{\mu},\;\phi_{\mu\nu}=\partial_{\mu}\phi_{\nu}-\partial_{\nu}\phi_{\mu}$, where $A^\mu$ is the photon field. The $\mathcal{L}_{\phi\gamma}$ we adopt here is a 
gauge invariant Lagrangian with the lowest derivative term. From Eq.(\ref{eq:vlllagrangian}), the general angular distribution of 
the leptonic decay can be obtained as
\begin{align}\label{eq:vllangledist}
\frac{1}{\Gamma}\dv{\Gamma}{\Omega}&=\frac{3}{16\pi}\bigg(2(1-\rho_{00})\cos^{2}\theta+(1+\rho_{00})\sin^{2}\theta+2\text{Re}[\rho_{1-1}]\sin^{2}\theta\cos2\phi-2\text{Im}[\rho_{1-1}]\sin^{2}\theta\sin2\phi\\
&+\sqrt{2}\text{Re}[\rho_{10}-\rho_{-10}]\sin2\theta\cos\phi-\sqrt{2}\text{Im}[\rho_{10}+\rho_{-10}]\sin2\theta\sin\phi\bigg).\nonumber
\end{align}
Similar to the $V\to PP$ decay, the normalized angular distribution is independent of the coupling constant 
and particle masses participating in the reaction.
Integration over $\phi$ again yields the $\theta$ distribution as
\begin{align}\label{eq:Wtheta2}
\frac{1}{\Gamma}\dv{\Gamma}{\cos\theta}&=\frac{3}{8}\left(1+\rho_{00}+(1-3\rho_{00})\cos^{2}\theta\right).
\end{align}
Note Eq.(\ref{eq:Wtheta2}) has a different $\theta$ dependence compared to Eq.(\ref{eq:Wtheta1}). Expressing this in terms of $\abs{\mathcal{M}}_T^2$ and $\abs{\mathcal{M}}_L^2$ given in Eq.(\ref{eq:tlamplitude}), one can rewrite this in the universal form given in Eq.(\ref{eq:polarangdist}). This difference compared to the $V\to PP$ decay leads to the different angular dependence of the transverse and longitudinal modes as shown in Fig.(\ref{fig:dilepton_cm_costheta_distribution}) for the $V\to ll$ decay. 
In particular, the longitudinal amplitude of the leptonic decay vanishes in the forward ($\theta = 0$) or backward ($\theta = \pi$) direction, whereas the transverse amplitude vanishes in the forward or backward direction for the $V\to PP$ decay. 
The difference arises from the way the angular momentum of the initial vector-meson is transferred to the final particles. 
As we can see in the Eq.(\ref{eq:vlllagrangian}), the $\phi$ meson couples to dileptons through a photon. Since helicity is conserved for (approximately) 
massless dileptons, only dileptons of opposite helicity are allowed. Thus in the forward or backward direction of the leptonic decay, the spins of the dileptons must be either aligned along forward or backward direction, and hence will not couple to the initial 
longitudinal $\phi$ meson mode. 
In Appendix \ref{appendix:four_vector_current}, we present additional calculations using explicit electron and positron spinors to 
derive the same result.

From an intuitive physics point of view, the difference in the angular dependencies of the  $V\to PP$ and $V\to ll$ 
decays comes from what carries the initial spin angular momentum of the vector meson. For the former, it is carried by the angular momentum, while for the latter the electron spin carries it. 
From a technical point of view, the difference is related to the different tensor structures shown in Eqs.(\ref{eq:vpplagrangian}) and (\ref{eq:vlllagrangian}). For the $V\to ll$ decay, the relative strength between the different tensor structures in the 
second expression of Eq.(\ref{eq:decaytensor}), which is controlled by the photon-electron coupling, determines the angular distribution. 
Moreover, it is uniquely fixed by the requirement that it vanishes when contracted with both $q^\mu$ and $k^\mu=(p_1^\mu-p_2^\mu)$. 
The first constraint holds exactly, as it is required from a Ward identity. 
However the second constraint is valid only to leading order in $\alpha$. 
At higher orders, a correction of the  $\sigma_{\mu\nu}/2m$ type, which is associated with the anomalous magnetic moment of the 
electron shows up.  
In case of the hadronic vertex $F_{\mu\nu} \phi^{\mu\nu}$ in Eq.(\ref{eq:vlllagrangian}), we again encounter higher-order terms with more (covariant) derivatives. However, by following the identical argument as for the previous $V\to PP$ decay, it can be shown that 
covariant derivative terms can only alter the overall constant of the decay amplitude, which can be absorbed into the definition of $g_J$. 
Therefore, the angular distribution of the leptonic decay remains intact.

\subsection{$A\to VP$}\label{subsection:k1decayrate}

For the $A\to VP$ decay, we study the $K_{1}\to\rho K$ and $K_{1}\to K^{*}\pi$ channels, for which the partial decay widths are taken from the PDG \cite{Workman:2022ynf} as

\begin{align}
\Gamma_{K_{1}\rho K}&=\frac{g_{K_{1}\rho K}^{2}}{8\pi}\abs{\bm{p_{1}}}\left(3+\frac{\abs{\bm{p_{1}}}^{2}}{m_{\rho}^{2}}\right)=34.2\,\text{MeV},\nonumber\\
&\\
\Gamma_{K_{1}K^{*}\pi}&=\frac{g_{K_{1}K^{*}\pi}^{2}}{8\pi}\abs{\bm{p_{1}}}\left(3+\frac{\abs{\bm{p_{1}}}^{2}}{m_{K^{*}}^{2}}\right)=18.9\,\text{MeV}.\nonumber
\end{align}

\begin{figure}[H]
\centering
\includegraphics[width=3.4in,height=1.6in]{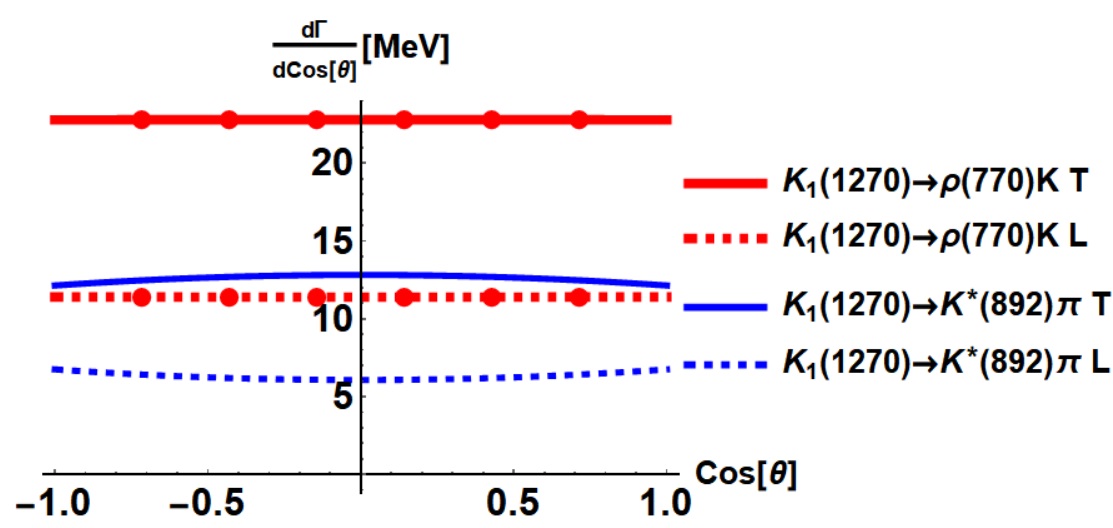}
\label{fig:rhokkstarpi}
\caption{Angular distribution of the decay rate of the $K_{1}\to\rho K$ and $K_{1}\to K^{*}\pi$ channels in the c.m. frame for each polarization. $T$ and $L$ represent the initial polarization.}
\label{fig:cos_theta_distribution2}
\end{figure}

The Lagrangian specifying the interaction between an axial vector meson, a vector meson and a pseudoscalar meson \cite{Sung:2021myr} is given as
\begin{align}\label{eq:k1lagrangian}
\mathcal{L}_{K_{1}VP}&=\sqrt{2}m_{K_{1}}\big(g_{K_{1}\rho K}\bar{K}\vec{\tau}\vdot\vec{\rho}_{\mu}-g_{K_{1}K^{*}\pi}\bar{K}^{*}_{\mu}\vec{\tau}\vdot\vec{\pi}\big)K_{1}^{\mu}.
\end{align}
Here, we take $m_{K_1}=1270\;\text{MeV}$. $g_{K_{1}\rho K}$ and $g_{K_{1}K^{*}\pi}$ are the coupling constants of $K_{1}\to\rho K$ and $K_{1}\to K^{*}\pi$, respectively given in Table \ref{table:moment_coupl}. 
The matrix representation of the $K_{1}$ field is given in Appendix \ref{appendix:effectivelagrangian}. 
We repeat the same procedure as in the two previous subsections, finally obtaining the general angular distribution of the $K_1\to VP$ decay as 
\begin{align}\label{eq:avpangledist}
\frac{1}{\Gamma}\dv{\Gamma}{\Omega}&=\frac{3}{4\pi\left(3+\frac{\abs{\bm{p_{1}}}^{2}}{m_{1}^{2}}\right)}\Bigg(1+\frac{\abs{\bm{p_{1}}}^{2}}{2m_{1}^{2}}\bigg(1-\rho_{00}+(3\rho_{00}-1)\cos^{2}\theta-2\text{Re}[\rho_{1-1}]\sin^{2}\theta\cos2\phi+2\text{Im}[\rho_{1-1}]\sin^{2}\theta\sin2\phi\\
&-\sqrt{2}\text{Re}[\rho_{10}-\rho_{-10}]\sin2\theta\cos\phi+\sqrt{2}\text{Im}[\rho_{10}+\rho_{-10}]\sin2\theta\sin\phi\bigg)\Bigg).\nonumber
\end{align}
Here, $m_{1}$ is the mass of the produced vector meson. We evaluate the polar angular distribution for $K_{1}\to\rho K(K^{*}\pi)$ by integrating over $\phi$, which gives
\begin{align}\label{eq:Wtheta3}
\frac{1}{\Gamma}\dv{\Gamma}{\cos\theta}&=\frac{3}{4\pi\left(3+\frac{\abs{\bm{p_{1}}}^{2}}{m_{1}^{2}}\right)}\left(1+\frac{\abs{\bm{p_{1}}}^{2}}{2m_{1}^2}\left(1-\rho_{00}+(3\rho_{00}-1)\cos^{2}\theta\right)\right).
\end{align}
This leads to,
\begin{align}
\dv{\Gamma}{\cos\theta}&=\frac{1}{16\pi}\frac{\abs{\bm{p_1}}}{E_{\text{c.m.}}^{2}}\left((1-\rho_{00})\frac{1}{2}\abs{\mathcal{M}_{TT}}^{2}+\rho_{00}\abs{\mathcal{M}_{LT}}^{2}+(1-\rho_{00})\frac{1}{2}\abs{\mathcal{M}_{TL}}^{2}+\rho_{00}\abs{\mathcal{M}_{LL}}^{2}\right)\\
&=\frac{1}{16\pi}\frac{\abs{\bm{p_1}}}{E_{\text{c.m.}}^{2}}2m_{K_{1}}^{2}g_{K_{1}VP}^{2}\left((1-\rho_{00})\left(1+\frac{\abs{\bm{p_{1}}}^{2}}{2m_{K^*}^{2}}\sin^{2}\theta\right)+\rho_{00}\left(1+\frac{\abs{\bm{p_{1}}}^{2}}{m_{K^*}^{2}}\cos^{2}\theta\right)\right).\nonumber
\end{align}
The two subscripts of $\mathcal{M}$ ($T$ or $L$) denote the polarization of the initial $K_1$ meson and final vector meson, respectively. Again, multiplying $\Gamma$ on both sides in the equation above, one obtains the equation in Eq.(\ref{eq:polarangdist}) with $\abs{\mathcal{M}}_{T}^{2}=\frac{1}{2}(\abs{\mathcal{M}_{TT}}^2+\abs{\mathcal{M}_{TL}}^2)$ and $\abs{\mathcal{M}}_{L}^2=\abs{\mathcal{M}_{LT}}^2+\abs{\mathcal{M}_{LL}}^2$. The factor of $1/2$ attached to $\abs{M}_{T}^2$ comes from the difference in degrees of freedom between transverse and longitudinal polarization of the initial state.
There is one crucial difference in Eq.(\ref{eq:Wtheta3}) compared to Eqs.(\ref{eq:Wtheta1}) and (\ref{eq:Wtheta2}), namely the existence of an extra momentum dependence in the second term inside the large bracket of the Eq.(\ref{eq:Wtheta3}). 
The resulting polar angular distributions of $K_{1}\to\rho K$ and $K_{1}\to K^{*}\pi$ are depicted in Fig.(\ref{fig:cos_theta_distribution2}). 
Unlike the case displayed in Figs.(\ref{fig:cos_theta_distribution}) and (\ref{fig:dilepton_cm_costheta_distribution}), 
it is difficult to discern $T$ and $L$ modes of the initial $K_{1}$ meson due to the abovementioned suppression factor $\abs{\bm{p_{1}}}^{2}/(2m_{1}^{2}) = 6\times10^{-4}$ ($\rho K),=5.6\times10^{-2}$ ($K^{*} \pi$) in the second term of the large bracket in Eq.(\ref{eq:Wtheta3}).

\begin{figure}[H]
\centering
\subfigure [$\dv{\Gamma}{\cos\theta}$ of $K_{1}\to\rho K$ in the c.m. frame]{
\includegraphics[width=3.1in,height=1.5in]{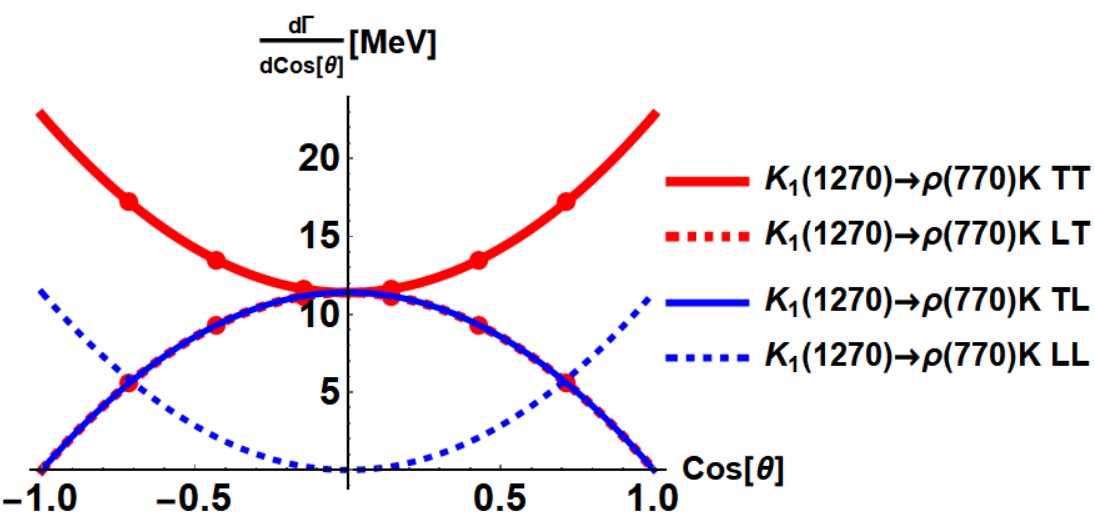}
\label{fig:k1rhokdecay}
}\subfigure [$\dv{\Gamma}{\cos\theta}$ of $K_{1}\to K^{*}\pi$ in the c.m. frame]{
\includegraphics[width=3.1in,height=1.5in]{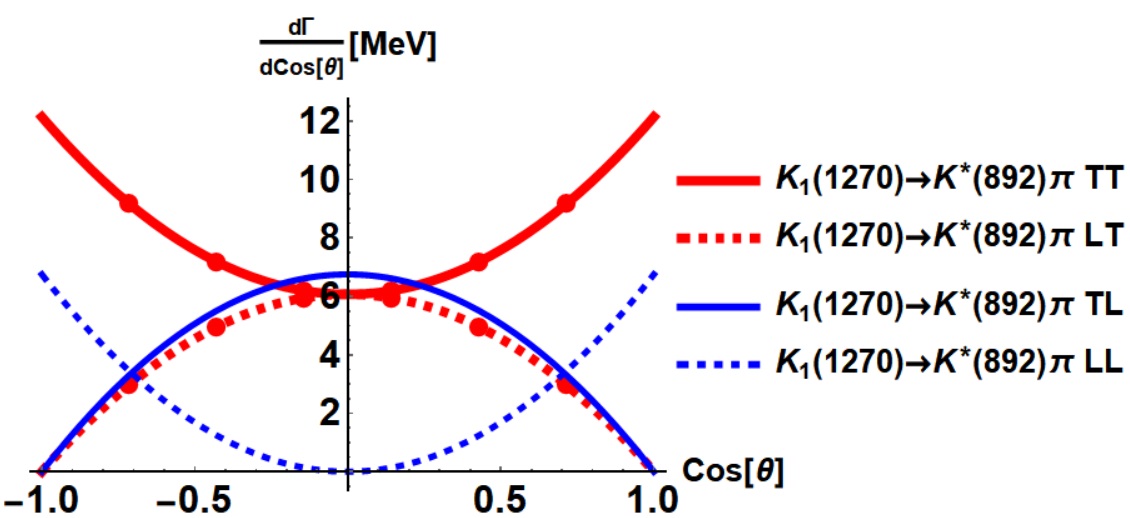}
\label{fig:k1kstarpidecay}
}
\caption{Angular distribution of the decay rate of (a) $K_{1}\to\rho K$ and (b) $K_{1}\to K^{*}\pi$ in the c.m. frame for each polarization. $T$ or $L$ appearing in the insert are the polarization of the initial $K_1$ meson and final vector-meson, respectively.}
\label{fig:cos_theta_distribution3}
\end{figure}

One can, however, distinguish the initial polarization by measuring the final vector-meson polarization, 
leading to a total of four possible combinations of initial axial vector meson and final vector meson polarizations. 
Explicit calculations of the corresponding amplitudes using the polarization vectors are given in Appendix \ref{appendix:polarizationvector}. The results of these calculations are given below.

\begin{align}
\abs{\mathcal{M}_{TT}}^{2}&=2m_{K_{1}}^{2}g_{K_{1}VP}^{2}(1+\cos^{2}\theta),\nonumber\\
\abs{\mathcal{M}_{LT}}^{2}&=2m_{K_{1}}^{2}g_{K_{1}VP}^{2}\sin^{2}\theta,\nonumber\\
\abs{\mathcal{M}_{TL}}^{2}&=2m_{K_{1}}^{2}g_{K_{1}VP}^{2}\frac{E_{1}^{2}}{m_{1}^{2}}\sin^{2}\theta,\\
\abs{\mathcal{M}_{LL}}^{2}&=2m_{K_{1}}^{2}g_{K_{1}VP}^{2}\frac{E_{1}^{2}}{m_{1}^{2}}\cos^{2}\theta.\nonumber\\
\end{align}
In Fig.\ref{fig:cos_theta_distribution3}, we display the above results. 
If the final vector meson is determined to be transversely polarized, we are able to discern the $T$ and $L$ mode of the initial $K_{1}$ meson by either observing the forward or backward directions. The same applies to the longitudinally polarized final vector meson. 
We can isolate either $T$ or $L$ modes of the initial $K_{1}$ meson by again observing the forward or backward direction. 

\section{Unified treatment using the Wigner $D$-matrix}\label{sec:wigmatrix}

In the previous section, we have computed the general angular distribution by using the corresponding interaction Lagrangian for each decay process. In this section, we are going to derive the same result by making use of the helicity formalism. 
In the 2-body decay helicity formalism, the quantization axis initially lies along the momentum of the initial particle in the Lab frame. 
The quantization axis is defined such that it diagonalizes the spin density matrix. 
We then apply a rotation to the quantization axis so that it aligns with one of the outgoing particle momentum in the initial particle rest frame. 
More details on the helicity formalism are discussed in Appendix \ref{appendix:helicityformalism}. The general angular distribution of a
spin-1 particle decay can be written as 
\begin{align}
\left(\frac{1}{\Gamma}\dv{\Gamma}{\Omega}\right)=\frac{1}{\Gamma}\frac{3}{4\pi}\sum_{\lambda_1,\lambda_2}\sum_{m,m^{\prime}=\pm1,0}\mathcal{D}^{1\dagger}(\phi,\theta,0)_{\lambda m}\rho_{m m^{\prime}}\mathcal{D}^{1}(\phi,\theta,0)_{m^{\prime}\lambda}\abs{H_{\text{int}}(\lambda_{1},\lambda_{2})}^{2}.
\label{eq:Wigner_D_Matrix}
\end{align}
Here, $\lambda=\lambda_{1}-\lambda_{2}$, where $\lambda_{1}$ and $\lambda_{2}$ stand for the helicities of the final particles with 
masses $m_{1}$ and $m_{2}$, respectively. 
The term $H(\lambda_{1},\lambda_{2})$ denotes the interaction Hamiltonian associated with each helicity component of the corresponding decay, which can be computed from the interaction Lagrangians given earlier. The convention of the Wigner $D$-matrix for a spin-1 particle is 
adopted from \cite{Wigner1959} and given as 
\begin{align}
&\mathcal{D}^{1}(\phi,\theta,0)_{mm^{\prime}}=\begin{pmatrix} \frac{1+\cos\theta}{2}e^{-i\phi} & -\frac{1}{\sqrt{2}}\sin\theta e^{-i\phi} & \frac{1-\cos\theta}{2}e^{-i\phi} \\ \frac{1}{\sqrt{2}}\sin\theta & \cos\theta & -\frac{1}{\sqrt{2}}\sin\theta \\ \frac{1-\cos\theta}{2}e^{i\phi} & \frac{1}{\sqrt{2}}\sin\theta e^{i\phi} & \frac{1+\cos\theta}{2} e^{i\phi}\end{pmatrix},
\label{eq:spin1rotation_matrix}
\end{align}
For the $V\to PP$ decay, $\lambda_{1}=\lambda_{2}=\lambda=0$. 
The transverse mode of the initial vector meson couples to a psuedoscalar meson decaying in the perpendicular direction. 
Also, the contribution comes only from the 00 element of the rotated spin density matrix. The general angular distribution of the $V\to PP$ 
decay is given as
\begin{align}
\left(\frac{1}{\Gamma}\dv{\Gamma}{\Omega}\right)_{PP} \propto \frac{1}{\Gamma}\frac{3}{4\pi}\sum_{m,m^{\prime}=\pm1,0}D^{1\dagger}_{0 m}\rho_{m m^{\prime}}D^{1}_{m^{\prime}0}. 
\end{align}
For the $V\to ll$ decay, the transversely polarized $\phi$ meson in contrast couples to a forwardly decaying dilepton. 
Also, by parity conservation in the electromagnetic decay, $\lambda = 1$ and $\lambda = -1$ contribute with equal weight. The general angular distribution is hence given as 
\begin{align}
\Biggl(\frac{1}{\Gamma}\dv{\Gamma}{\Omega}\Biggr)_{e^+e^-} \propto\frac{1}{\Gamma}\frac{3}{4\pi} \sum_{\lambda = \pm 1} \sum_{m,m^{\prime}=\pm1,0}D^{1\dagger}_{\lambda m}\rho_{m m^{\prime}}D^{1}_{m^{\prime} \lambda}.
\end{align}
For the $A\to VP$ decay, $\lambda=\pm1,0$ contribute to the distribution, which is calculated to be 
\begin{align}
\left(\frac{1}{\Gamma}\dv{\Gamma}{\Omega}\right)_{VP} \propto\frac{1}{\Gamma}\frac{3}{4\pi}\left(\sum_{\lambda=\pm1}\sum_{m,m^{\prime}=\pm1,0}D^{1\dagger}_{\lambda m}\rho_{m m^{\prime}}D^{1}_{m^{\prime}\lambda}\abs{H_{\text{int}}(1,0)}^{2}+\sum_{m,m^{\prime}=\pm1,0}D^{1\dagger}_{0m}\rho_{m m^{\prime}}D^{1}_{m^{\prime}0}\abs{H_{\text{int}}(0,0)}^{2}\right). 
\end{align}
Using the above formulas, we can reproduce all the results given in Eq.(\ref{eq:vppangledist}), (\ref{eq:vllangledist}) and (\ref{eq:avpangledist}).

\section{Summary and Conclusions}\label{sec:sumandconc}
In a nuclear medium, a spin-1 particle moving relative to the medium reacts differently depending on its polarization direction. A longitudinally polarized mode is a state in which the spin of the spin-1 meson points in the perpendicular direction of its motion, while a transversely polarized mode occurs when its spin points in the parallel directions. 
The different responses of the two modes manifest as two separate peaks in the invariant mass spectrum, mimicking a broadened width when hadrons with various momenta are observed. 
The two modes become degenerate when the particle is at rest with respect to the medium. To identify a possible momentum-independent mass shift caused by chiral symmetry restoration in the nuclear medium, it is therefore important to distinguish between the two modes.

In this paper, we have discussed how the polarization modes of a massive spin-1 meson can be disentangled in its two body decay by measuring the decay angles and the polarization of a produced particle. 
This was shown by computing the general angular distribution of the two body decay from the interaction Lagrangian, the result of which can also be derived by making use of helicity formalism. 
For then $V\to PP$ channel, the forward or backward angle is dominated by a longitudinally polarized vector meson. On the contrary, the forward or backward angle in the $V\to ll$ decay is dominated by a transversely polarized vector meson. 
Both measurements in nuclear target experiments have pros and cons. 
While the signal from the former reaction will be distorted by the final state interaction, it is possible to perform 
high statistics measurement due to its large branching ratio. 
For the latter reaction, the electromagnetic signal will not be distorted by the surrounding matter, while its branching ratio is very small.  In J-PARC experiments both measurements are planned for the $\phi$ meson, E88 for the former and E16 for the latter. 
For the $e^{+}e^{-}$ channel, the E16 experiment has a small acceptance for low momentum leptons 
%with a momentum lower than $0.4 \text{GeV}/c$
%"90\% efficiency for 0.4~GeV."
since it requires a minimum energy at the trigger level in order to suppress the background and low momentum leptons have a large bending angle in the magnetic field of the E16 spectrometer. In particular, this will not be favorable for measuring the backward emitted leptons which will have a small momentum due to the Lorentz boost to the Lab frame. As for the $K^{+}K^{-}$ channel to be measured by the J-PARC E88 experiment, it is expected that due to the small Q-value and Lorentz boost resulting in a small opening angle in the Lab frame, an approximately full and uniform coverage of all decay angles will be possible. 
Therefore, an extraction of transversely-polarized $\phi$ mesons at $\theta=90^{\circ}$ and longitudinally polarized ones at $\theta=0,\;\pi$ is expected to be possible.  
Therefore, the two experiments will be complementary to each other and provide important information needed for the final reconstruction of the possible mass shift of the $\phi$ meson in nuclear matter.

As for the  $A\to VP$ process, the decay angle distribution alone turns out to be insufficient to distinguish the polarization of the initial axial vector meson. However, as we have demonstrated in this paper, 
this matter can be resolved by measuring the polarization of the final vector meson. 
Before long, such a measurement will be available in a future J-PARC experiment, which will aid in reducing the uncertainty of the $K^*$ and $K_1$ mass shifts in nuclear matter. Once this is achieved, the result will provide a deeper understanding of how chiral symmetry breaking (and its restoration) contributes to hadron mass generation

\section*{Acknowledgments}
This work was supported by the Korea National Research Foundation under grant No. 2023R1A2C3003023 and No. 2023K2A9A1A0609492411, and by JSPS KAKENHI Grants No. JP19KK0077, No. JP20K03940, No. JP21H00128, No. JP21H00128, and No. JP21H01102, 
for the Promotion of Science (JSPS). This work was also supported by the JAEA REIMEI project under the title ``Studying the origin of hadron masses through the behavior of vector mesons in nuclear matter from theory and experiment".

\clearpage
\appendix

\renewcommand{\thesection}{\Alph{section}}
\renewcommand{\thesubsection}{\Alph{section}-\arabic{subsection}}

\section{Effective interaction Lagrangian}\label{appendix:effectivelagrangian}

In this section, we introduce the effective Lagrangian for each decay process. Each Lagrangian is assigned to one of three different categories: a vector meson decaying into two pseudoscalar mesons ($V\to PP$), a vector meson decaying into dileptons ($V\to ll$) and an axial vector meson decaying into a vector meson and a pseudoscalar meson ($A\to VP$). 
The $K$, $K^{*}$ and $K_{1}$ isodoublet matrices are defined as follows,
\begin{align}
K^{*}&=\begin{pmatrix}K^{+*}\\K^{0*}\end{pmatrix},\;\bar{K}=\begin{pmatrix}K^{-}&\bar{K}^{0}\end{pmatrix},\,K_{1}=\begin{pmatrix}K_{1}^{+}\\K_{1}^{0}\end{pmatrix},\,\bar{K}^{*}=\begin{pmatrix}K^{-*}&\bar{K}^{0*}\end{pmatrix}.
\end{align}
The $\pi$ and $\rho$ meson isotriplet fields are defined as follows along with the Pauli matrix $\bm{\tau}$, 
\begin{align}
(\pi/\rho)^{0}&=(\pi/\rho)^{3},\;(\pi/\rho)^{\pm}=\frac{1}{\sqrt{2}}\left((\pi/\rho)^{1}\mp i(\pi/\rho)^{2}\right),\nonumber\\
&\\
\tau_{1}&=\begin{pmatrix}0&1\\1&0\end{pmatrix},\;\tau_{2}=\begin{pmatrix}0&-i\\i&0\end{pmatrix},\;\tau_{3}=\begin{pmatrix}1&0\\0&-1\end{pmatrix}.\nonumber
\end{align}
$\phi\to K^+K^-,\;\rho\to\pi\pi$ and $K^{*}\to K\pi$ are assigned to the $V\to PP$ category, and their interaction Lagrangians can be given as

\begin{align}
\mathcal{L}_{\phi K^+K^-}&=g_{\phi K^+K^-}(K^{+}\partial_{\mu}K^{-}-K^{-}\partial_{\mu}K^{+})\phi^{\mu},\nonumber\\
\mathcal{L}_{\rho\pi\pi}&=g_{\rho\pi\pi}\bigg((\pi^{+}\partial_{\mu}\pi^{-}-\partial_{\mu}\pi^{+}\pi^{-})\rho^{0\mu}+(\pi^{-}\partial_{\mu}\pi^{0}-\partial_{\mu}\pi^{-}\pi^{0})\rho^{+\mu}+(\pi^{+}\partial_{\mu}\pi^{0}-\partial_{\mu}\pi^{+}\pi^{0})\rho^{-\mu}\bigg),\\
\mathcal{L}_{K^{*}K\pi}&=g_{K^{*}K\pi}\bigg[\sqrt{2}\bigg(K^{-}\partial_{\mu}\pi^{0}-\partial_{\mu}K^{-}\pi^{0}+\bar{K}^{0}\partial_{\mu}\sqrt{2}\pi^{-}-\partial_{\mu}\bar{K}^{0}\sqrt{2}\pi^{-}\bigg)K^{+*\mu}\nonumber\\
&+\sqrt{2}\bigg(K^{-}\partial_{\mu}\sqrt{2}\pi^{+}-\partial_{\mu}K^{-}\sqrt{2}\pi^{+}-\bar{K}^{0}\partial_{\mu}\pi^{0}+\partial_{\mu}\bar{K}^{0}\pi^{0}\bigg)K^{0*\mu}\bigg].\nonumber
\end{align}
For the $\phi$ meson decaying into a positron and an electron, the $\phi$-photon coupling Lagrangian is given by the vector meson dominance (VMD) model, as 
\begin{align}
\mathcal{L}_{\phi\gamma}&=-\frac{e}{2g_{J}}F^{\mu\nu}\phi_{\mu\nu},\;\mathcal{L}_{\gamma e^+e^-}=-e\bar{\psi}\gamma^{\mu}A_{\mu}\psi.
\end{align}
Finally, the Lagrangian for a $K_{1}$ decaying into a vector meson and a pseudoscalar meson is given by a direct matrix multiplication as
\begin{align}
\mathcal{L}_{K_{1}VP}&=m_{K_{1}}g_{K_{1}K^{*}\pi}\sqrt{2}\bigg[\bigg(K^{-*}_{\mu}\pi^{0}+\bar{K}^{0*}_{\mu}\sqrt{2}\pi^{-}\bigg)K_{1}^{+\mu}+\bigg(K^{-*}_{\mu}\sqrt{2}\pi^{+}-\bar{K}^{0*}_{\mu}\pi^{0}\bigg)K_{1}^{0\mu}\bigg]\\
&-m_{K_{1}}g_{K_{1}\rho K}\sqrt{2}\bigg[\bigg(K^{-}\rho^{0}_{\mu}+\bar{K}^{0}\sqrt{2}\rho^{+}_{\mu}\bigg)K_{1}^{+\mu}+\bigg(K^{-}\sqrt{2}\rho^{+}_{\mu}-\bar{K}^{0}\rho^{0}_{\mu}\bigg)K_{1}^{0\mu}\bigg].\nonumber
\end{align}

\section{Spin and polarization of a massive spin 1 particle}\label{appendix:basis}

Our calculation of the angular distribution is carried out by making use of the circularly polarized bases of a
massive spin-1 particle. Such bases consist of eigenvectors of $J_3$ of the SO(3) group. The 3 generators of the SO(3) group can be derived from a general rotation matrix in 3D space, written in terms of a rotation angle $\theta$ and rotation axis $\hat{n}=(n_{1},n_{2},n_{3})$ (normalized to unity).

\begin{align}\label{eq:rotmatrix}
R(\hat{n},\theta)=\begin{pmatrix}\cos\theta+(1-\cos\theta)n_{1}^{2}&(1-\cos\theta)n_{1}n_{2}-n_{3}\sin\theta &(1-\cos\theta)n_{1}n_{3}+n_{2}\sin\theta \\(1-\cos\theta)n_{1}n_{2}+n_{3}\sin\theta &\cos\theta+(1-\cos\theta)n_{2}^{2}&(1-\cos\theta)n_{2}n_{3}-n_{1}\sin\theta \\(1-\cos\theta)n_{1}n_{3}-n_{2}\sin\theta &(1-\cos\theta)n_{2}n_{3}+n_{1}\sin\theta &\cos\theta+(1-\cos\theta)n_{3}^{2}\end{pmatrix}.
\end{align}
From the above rotation matrix, we can derive rotation generators about each axis as 
\begin{align}
\label{eq:rotgen}
&J_{1}=\begin{pmatrix}0&0&0\\0&0&-i\\0&i&0\end{pmatrix},\;J_{2}=\begin{pmatrix}0&0&i\\0&0&0\\-i&0&0\end{pmatrix},\;J_{3}=\begin{pmatrix}0&-i&0\\i&0&0\\0&0&0\end{pmatrix},
\end{align}
where $(J_{i})_{jk}=-i\epsilon_{ijk}$ and $\comm{J_{i}}{J_{j}}=i\epsilon_{ijk}J_{k}$. 1,2, and 3 correspond to $x$,$y$ and $z$. 
The three eigenvectors of $J_{3}$ can be used to construct the spatial part of the circularly polarized bases 
in the rest frame of a massive spin-1 particle. 
$\bm{\varepsilon_{\pm1}}$ are the transversely polarized vectors with eigenvalues $J_{3}=\pm1$, respectively and $\bm{\varepsilon_{0}}$ 
is the longitudinally polarized vector with eigenvalue $J_{3}=0$. 
The corresponding states and four-momentum of the initial massive spin-1 particle are given as 
\begin{align}
\label{eq:state}
&\bm{\varepsilon_{\pm1}}=\left(\mp\frac{1}{\sqrt{2}},-\frac{i}{\sqrt{2}},0\right),\;\bm{\varepsilon_{0}}=\left(0,0,1\right),\nonumber\\
&\ket{\pm1}=\left(0,\bm{\varepsilon_{\pm1}}\right),\;\ket{0}=\left(0,\bm{\varepsilon_{0}}\right),\\
&\ket{V}=\sum_{\lambda=\pm1,0}a_{\lambda}\ket{\lambda},\;q^{\mu}=\left(m_{0},0, 0,0\right).\nonumber
\end{align}

\section{Decay amplitude computation and polar angular dependence in the initial particle rest frame}\label{appendix:angular_dep_comp}

In this appendix, we will evaluate the polar angle decay distribution of each polarization mode of a massive spin-1 particle using its polarization tensor. Decay processes to be considered are $V\to PP$, $V\to ll$ and $A\to VP$. 

\subsection{Polarization modes of a massive spin-1 particle}\label{appendix:polarization_tensor}
Let's begin with the simple case for the decay of a purely transversely or longitudinally-polarized state. We can construct the projection operator using polarization vectors given in Appendix B. They are acquired as 
\begin{align}\label{eq:projection_op}
&P_{T}^{\mu\nu}=\sum_{\lambda=\pm1}\varepsilon_{\lambda}^{\mu}\varepsilon_{\lambda}^{\nu*}=\begin{pmatrix}0&0\\0&\delta^{ij}-\frac{q^{i}q^{j}}{\bm{q}^{2}}\end{pmatrix},\nonumber\\
&\\
&P_{L}^{\mu\nu}=\varepsilon_{0}^{\mu}\varepsilon_{0}^{\nu*}=\begin{pmatrix}\frac{\bm{q}^{2}}{m_{0}^{2}}&\frac{q_{0}q^{i}}{m_{0}^{2}}\\\frac{q_{0}q^{i}}{m_{0}^{2}}&\frac{q_{0}^{2}q^{i}q^{j}}{m_{0}^{2}\bm{q}^{2}}\nonumber
\end{pmatrix}.
\end{align}
The subscripts $T$ and $L$ denote the transverse and longitudinal polarization modes, respectively. 
$q_{0}$ and $q^{i}$ are the energy and momentum of the spin-1 particle, where $i$ and $j$ run from 1 to 3 and $q^{i}=0$ in the initial particle rest frame. 
To calculate the decay amplitude, we need the corresponding matrix element from an effective Lagrangian. We will take $\phi\to K^+K^-$ as an example for the $V\to PP$ decay, $\phi\to e^{+}e^{-}$ for $V\to ll$ and finally $K_{1}\to K^{*}\pi$ for the $A\to VP$ decay. 
The same method can be applied to other decay processes. Note that for the $K$ meson decay one needs to consider isospin degrees of freedom. The matrix element for each decay process is given as 
\begin{align}\label{eq:decaytensor}
\mathcal{M}^{VPP}_{\mu\nu}&=g_{\phi K^{+}K^{-}}^{2}(p_{1}-p_{2})_{\mu}(p_{1}-p_{2})_{\nu},\nonumber\\
\mathcal{M}^{Vll}_{\mu\nu}&=\frac{64\pi^{2}\alpha^{2}}{g_{J}^{2}}\bigg(p_{1\mu}p_{2\nu}+p_{2\mu}p_{1\nu}-\frac{1}{2}m_{\phi}^{2}g_{\mu\nu}\bigg),\\
\mathcal{M}^{AVP}_{\mu\nu}&=2m_{K_{1}}^{2}g_{K_{1}K^{*}\pi}^{2}\varepsilon_{\mu}(\lambda_{K^*})\varepsilon_{\nu}^{*}(\lambda_{K^*}).\nonumber
\end{align}
Here, $p_{1}^{\mu} = (E_1, \bm{p}_1)$ and $p_{2}^{\mu} = (E_2, \bm{p}_2)$ are the four-momentum of a produced particles and $\lambda_{K^*}$ is the helicity of the $K^{*}$ meson. 
The decay amplitudes can be obtained by multiplying the projection operators of Eq.\,(\ref{eq:projection_op}) and contracting the Lorentz indices. The specific results are given as 
\begin{align}\label{eq:tlamplitude}
\phi\to K^{+}+K^{-}&\begin{cases}&|\mathcal{M}|_{T}^{2}=2g_{\phi K^+K^-}^{2}\abs{\bm{p_{1}}}^{2}\sin^{2}\theta,\\&\\&|\mathcal{M}|_{L}^{2}=4g_{\phi K^+K^-}^{2}\abs{\bm{p_{1}}}^{2}\cos^{2}\theta,\end{cases}\nonumber\\\nonumber\\
\phi\to e^{+}+e^{-}&\begin{cases}&|\mathcal{M}|_{T}^{2}=\frac{64\pi^{2}\alpha^{2}}{g_{J}^{2}}\frac{1}{4}m_{\phi}^{2}\left(1+\sin^{2}\theta\right),\\\\&|\mathcal{M}|_{L}^{2}=\frac{64\pi^{2}\alpha^{2}}{g_{J}^{2}}\frac{1}{2}m_{\phi}^{2}\sin^{2}\theta,\end{cases}\\\nonumber\\
K_{1}\to K^*+\pi&\begin{cases}&\abs{\mathcal{M}}_{T}^{2}=m_{K_{1}}^{2}g_{K_{1}K^*\pi}^{2}\left(2+\frac{\abs{\bm{p_{1}}}^{2}}{m_{K^*}^{2}}\sin^{2}\theta\right),\\&\\&\abs{\mathcal{M}}_{L}^{2}=2m_{K_{1}}^{2}g_{K_{1}K^*\pi}^{2}\left(1+\frac{\abs{\bm{p_{1}}}^{2}}{m_{K^*}^{2}}\cos^{2}\theta\right).\end{cases}\nonumber
\end{align}
The different $\abs{\bm{p_{1}}}^{2}$ values for each decay mode are given below.

\subsection{General angular distribution of the $\phi\to K^{+}+K^{-}$ decay}\label{appendix:kaon_distribution}
Next, we study the more general angular distribution, where the initial particle state is composed of a linear superposition of three different helicity states. From the interaction Lagrangian in Eq.(\ref{eq:vpplagrangian}), the invariant amplitude is written as 
\begin{align}\label{eq:vppamp}
\mathcal{M}_{\phi K^+K^-}&=g_{\phi K^+K^-}(p_{1}-p_{2})_{\mu}\sum_{\lambda_{\phi}=\pm1,0}a_{\lambda_{\phi}}\varepsilon^{\mu}(\lambda_{\phi}),
\end{align}
where $p_{1}^{\mu} = (E_1, \bm{p}_1)$ and $p_{2}^{\mu} = (E_2, \bm{p}_2)$ are the four-momenta of the produced kaons, with 
$\abs{\bm{p_1}} = \abs{\bm{p_2}}=\frac{1}{2}\sqrt{m_{\phi}^{2}-4m_{K}^{2}}$ in the initial $\phi$ meson rest frame, where $m_K$ stands for the kaon mass. 
By taking the absolute square, we can obtain the general angular distribution as detailed in Ref.\cite{Schilling:1969um}. The results are given as 
\begin{align}
\abs{\mathcal{M}_{\phi K^+K^-}}^{2}&=2g_{\phi K^+K^-}^{2} \abs{\bm{p}_1}^2 
\bigg(2\rho_{00}\cos^{2}\theta+(1-\rho_{00})\sin^{2}\theta-2\text{Re}[\rho_{1-1}]\sin^{2}\theta\cos2\phi+2\text{Im}[\rho_{1-1}]\sin^{2}\theta\sin2\phi\\
&-\sqrt{2}\text{Re}[\rho_{10}-\rho_{-10}]\sin2\theta\cos\phi+\sqrt{2}\text{Im}[\rho_{10}+\rho_{-10}]\sin2\theta\sin\phi\bigg).\nonumber
\end{align}

\subsection{General angular distribution of the $\phi\to e^{+}+e^{-}$ decay}\label{appendix:dilepton_distribution}
The $\phi\to\gamma^{*}\to e^{+}+e^{-}$ decay amplitude can be computed using the interaction Lagrangian given in Eq.(\ref{eq:vlllagrangian}), and is obtained as
\begin{align}
\mathcal{M}_{\phi e^{+}e^{-}}(\lambda_{1},\lambda_{2})&=\frac{4\pi\alpha}{g_{J}}\bar{u}^{\lambda_{2}}(p_{2})\gamma^{\mu}\sum_{\lambda_{\phi}=\pm1,0}a_{\lambda_{\phi}}\varepsilon_{\mu}(\lambda_{\phi})v^{\lambda_{1}}(p_{1}). 
\end{align}
$p_{1}^{\mu} = (E_1, \bm{p}_1)$ and $p_{2}^{\mu} = (E_2, \bm{p}_2)$ are the four-momenta of the produced positron and electron, respectively. $v^{\lambda_1}(p_{1})$ and $u^{\lambda_2}(p_{2})$ are the Dirac spinors of the positron with helicity $\lambda_{1}$ and electron with helicity $\lambda_{2}$, respectively. 
$\abs{\bm{p_1}}= \abs{\bm{p_2}}=\frac{1}{2}\sqrt{m_{\phi}^{2}-4m_{l}^{2}} \approx \frac{1}{2}m_{\phi}$, 
where $m_{l}$ is the electron mass, which is approximated as 0 here. 
We obtain the general angular distribution by taking the absolute square and summing over the final dilepton polarizations. 
\begin{align}
\sum_{\lambda_{1},\lambda_{2}=\pm1/2}\abs{\mathcal{M}_{\phi e^{+}e^{-}}(\lambda_{1},\lambda_{2})}^{2}&=\frac{16\pi^{2}\alpha^{2}m_{\phi}^{2}}{g_{J}^{2}}\bigg(1+\rho_{00}+(1-3\rho_{00})\cos^{2}\theta+2\text{Re}[\rho_{1-1}]\sin^{2}\theta\cos2\phi\\
&-2\text{Im}[\rho_{1-1}]\sin^{2}\theta\sin2\phi+\sqrt{2}\text{Re}[\rho_{10}-\rho_{-10}]\sin2\theta\cos\phi-\sqrt{2}\text{Im}[\rho_{10}+\rho_{-10}]\sin2\theta\sin\phi\bigg).\nonumber
\end{align}

\subsection{General angular distribution of the $K_{1}\to K^*+\pi$ decay}\label{appendix:k1angdist}
From the interaction Lagrangian given in the Eq.(\ref{eq:k1lagrangian}), the invariant amplitude is obtained as
\begin{align}
\mathcal{M}_{K_{1}K^*\pi}(\lambda_{K^*})&=\sqrt{2}m_{K_{1}}g_{K_{1}K^*\pi}\sum_{\lambda_{K_{1}}=\pm1,0}a_{\lambda_{K_{1}}}\varepsilon^{\mu}(\lambda_{K_{1}})\varepsilon^{*}_{\mu}(\lambda_{K^*}).
\end{align}
$\lambda_{K_{1}}$ and $\lambda_{K^*}$ denote the helicities of the $K_{1}$ and $K^*$ mesons. 
To calculate the scalar product, we need the polarization vector of the final vector meson in the initial axial vector meson rest frame which can be obtained by an Euler rotation $R(\phi,\theta,0)$ following a Lorentz boost along the $z$-axis. 
By multiplying the explicit Lorentz transformation matrix given in the Eq.(\ref{eq:helboost}) to the polarization vector given in Eq.(\ref{eq:state}), we obtain the polarization vectors of the final vector meson as 
\begin{align}
&\varepsilon^{\mu}(\bm{p_1},\pm1)=\begin{pmatrix}0\\\mp\frac{1}{\sqrt{2}}\cos\theta\cos\phi+\frac{i}{\sqrt{2}}\sin\phi\\\mp\frac{1}{\sqrt{2}}\cos\theta\sin\phi-\frac{i}{\sqrt{2}}\cos\phi\\\pm\frac{1}{\sqrt{2}}\sin\theta\end{pmatrix},\,\varepsilon^{\mu}(\bm{p_1},0)=\begin{pmatrix}\frac{\abs{\bm{p_1}}}{m_{1}}\\\frac{E_{1}}{m_{1}}\sin\theta\cos\phi\\\frac{E_{1}}{m_{1}}\sin\theta\sin\phi\\\frac{E_{1}}{m_{1}}\cos\theta\end{pmatrix},
\end{align}
where $E_{1}$ is the energy of the final vector meson in the c.m. frame. 
The dot product between the initial and final polarization vectors are computed as
\begin{align}
\varepsilon_{K_{1}}(1)\vdot\varepsilon_{K^*}^{*}(\pm1)&=-\frac{1\pm\cos\theta}{2}e^{i\phi},\,\varepsilon_{K_{1}}(-1)\vdot\varepsilon_{K^*}^{*}(\pm1)=-\frac{1\mp\cos\theta}{2}e^{-i\phi},\nonumber\\
\varepsilon_{K_{1}}(0)\vdot\varepsilon_{K^*}^{*}(\pm1)&=\mp\frac{1}{\sqrt{2}}\sin\theta,\;\;\qquad\varepsilon_{K_{1}}(\pm1)\vdot\varepsilon_{K^*}^{*}(0)=\pm\frac{E_{1}}{\sqrt{2}m_{1}}\sin\theta e^{\pm i\phi},\\
\varepsilon_{K_{1}}(0)\vdot\varepsilon_{K^*}^{*}(0)&=-\frac{E_{1}}{m_{1}}\cos\theta.\nonumber
\end{align}
Finally, the general angular distribution can be given as
\begin{align}
\sum_{\lambda_{K^*}=\pm1,0}\abs{\mathcal{M}_{K_{1}K^*\pi}(\lambda_{K^*})}^{2}&=2m_{K_{1}}^{2}g_{K_{1}K^*\pi}^{2}\Bigg(1+\frac{\abs{\bm{p_{1}}}^{2}}{2m_{1}^{2}}\bigg(1-\rho_{00}+(3\rho_{00}-1)\cos^{2}\theta-2\text{Re}[\rho_{1-1}]\sin^{2}\theta\cos2\phi\\
&+2\text{Im}[\rho_{1-1}]\sin^{2}\theta\sin2\phi-\sqrt{2}\text{Re}[\rho_{10}-\rho_{-10}]\sin2\theta\cos\phi+\sqrt{2}\text{Im}[\rho_{10}+\rho_{-10}]\sin2\theta\sin\phi\bigg)\Bigg)\nonumber.
\end{align}

\section{Vector current $\bar{u}(p_{2})\gamma^{\mu}v(p_{1})$ generating helicity states of dileptons}\label{appendix:four_vector_current}
The helicity structure of the produced positron and electron provides an intuitive picture of 
the $\phi\to e^{+}+e^{-}$ decay amplitude. 
In this subsection, we we discuss the explicit details of this point, by exploiting the properties of Dirac spinors in the $\phi$ meson rest frame. $p_{1},\;E_1$ and $p_{2},\;E_2$ are the four-momenta and energies of the outgoing positron and electron, respectively, where $\abs{\bm{p_1}}=\abs{\bm{p_2}}=\frac{1}{2}m_{\phi}$. 
Since the energies of positron and electron are identical, we denote them as $E_{1}=E_{2}=E$. 
As the outgoing dileptons are approximated to be massless, we adopt the spinors in the chiral representation, and thus have 
\begin{align}
&\gamma_{0}=\begin{pmatrix}0&1\\1&0\end{pmatrix},\;\gamma_{i}=\begin{pmatrix}0&\sigma_{i}\\-\sigma_{i}&0\end{pmatrix},\;\gamma_{5}=\begin{pmatrix}-1&0\\0&1\end{pmatrix}.
\end{align}

First, we study the case where the positron (electron) is emitted into the forward (backward) direction. 
This implies that their spinors are eigenstates of the spin operator with respect to the $z$-axis, which in the massless limit can be written as,
\begin{align}
&v^{1}(p_{1})=\sqrt{2E_{1}}\begin{pmatrix}0\\1\\0\\0\end{pmatrix},\;v^{2}(p_{1})=\sqrt{2E_{1}}\begin{pmatrix}0\\0\\-1\\0\end{pmatrix},\nonumber\\
&\\
&u^{1}(p_{2})=\sqrt{2E_{2}}\begin{pmatrix}1\\0\\0\\0\end{pmatrix},\;u^{2}(p_{2})=\sqrt{2E_{2}}\begin{pmatrix}0\\0\\0\\1\end{pmatrix}.\nonumber
\end{align}
The vector current for each helicity combination can be written as
\begin{align}
&\bar{u}^{1}\gamma^{\mu}v^{1}=\begin{pmatrix}0,&-2E,&2iE,&0\end{pmatrix},\;\bar{u}^{1}\gamma^{\mu}v^{2}=\begin{pmatrix}0,&0,&0,&0\end{pmatrix},\nonumber\\
&\\
&\bar{u}^{2}\gamma^{\mu}v^{1}=\begin{pmatrix}0,&0,&0,&0\end{pmatrix},\;\bar{u}^{2}\gamma^{\mu}v^{2}=\begin{pmatrix}0,&-2E,&-2iE,&0\end{pmatrix}.\nonumber
\end{align} 
The vector current going through the general angle $(\phi,\theta)$ can be obtained by applying an Euler rotation $R(\phi,\theta,0)$.
\begin{align}
&\begin{pmatrix}1&0&0&0\\0&\cos\phi&-\sin\phi&0\\0&\sin\phi&\cos\phi&0\\0&0&0&1\end{pmatrix} \cdot \begin{pmatrix}1&0&0&0\\0&\cos\theta&0&\sin\theta\\0&0&1&0\\0&-\sin\theta&0&\cos\theta 
\end{pmatrix},
\end{align}
which yields the nonvanishing vector currents with opposite helicity:
\begin{align}
&\bar{u}^{1}(p_{2})\gamma^{\mu}v^{1}(p_{1})=2E\begin{pmatrix}0&-\cos\theta\cos\phi-i\sin\phi&-\cos\theta\sin\phi+i\cos\phi&\sin\theta\end{pmatrix},\nonumber\\
&\\
&{u}^{2}(p_{2})\gamma^{\mu}v^{2}(p_{1})=2E\begin{pmatrix}0&-\cos\theta\cos\phi+i\sin\phi&-\cos\theta\sin\phi-i\cos\phi&\sin\theta\end{pmatrix}.\nonumber
\end{align}

\section{Final vector meson polarization measurement in the $A\to VP$ decay}\label{appendix:polarizationvector}

As we have studied in Appendix \ref{appendix:k1angdist}, we know the explicit form of the scalar product between the polarization vectors of the initial axial vector meson and the final vector meson in the $A\to VP$ decay. 
Since the decay amplitudes for the transverse and longitudinal modes of the initial $K_{1}$ are (practically) indistinguishable, we aim to resolve this matter by measuring the final vector meson polarization. 
Additionally, using the scalar product of the polarization vectors of the initial axial vector meson and the final vector meson, we obtain the four possible amplitudes depicted in Fig.\ref{fig:cos_theta_distribution3} as 
\begin{align}
\abs{\mathcal{M}_{TT}}^{2}&=2m_{K_{1}}^{2}g_{K_{1}VP}^{2}\sum_{\lambda_{K_{1}}=\pm1}\sum_{\lambda_{V}=\pm1}\abs{\varepsilon^{\mu}(\lambda_{K_{1}})\varepsilon_{\mu}^{*}(\lambda_{V})}^{2}=2m_{K_{1}}^{2}g_{K_{1}VP}^{2}(1+\cos^{2}\theta),\nonumber\\
\abs{\mathcal{M}_{LT}}^{2}&=2m_{K_{1}}^{2}g_{K_{1}VP}^{2}\sum_{\lambda_{K_{1}}=0}\sum_{\lambda_{V}=\pm1}\abs{\varepsilon^{\mu}(\lambda_{K_{1}})\varepsilon_{\mu}^{*}(\lambda_{V})}^{2}=2m_{K_{1}}^{2}g_{K_{1}VP}^{2}\sin^{2}\theta,\nonumber\\
\abs{\mathcal{M}_{TL}}^{2}&=2m_{K_{1}}^{2}g_{K_{1}VP}^{2}\sum_{\lambda_{K_{1}}=\pm1}\sum_{\lambda_{V}=0}\abs{\varepsilon^{\mu}(\lambda_{K_{1}})\varepsilon_{\mu}^{*}(\lambda_{V})}^{2}=2m_{K_{1}}^{2}g_{K_{1}VP}^{2}\frac{E_{1}^{2}}{m_{1}^{2}}\sin^{2}\theta,\\
\abs{\mathcal{M}_{LL}}^{2}&=2m_{K_{1}}^{2}g_{K_{1}VP}^{2}\sum_{\lambda_{K_{1}}=0}\sum_{\lambda_{V}=0}\abs{\varepsilon^{\mu}(\lambda_{K_{1}})\varepsilon_{\mu}^{*}(\lambda_{V})}^{2}=2m_{K_{1}}^{2}g_{K_{1}VP}^{2}\frac{E_{1}^{2}}{m_{1}^{2}}\cos^{2}\theta.\nonumber
\end{align}

\section{More details about the helicity formalism}\label{appendix:helicityformalism}
The two-body decay process begins from a particle state of definite angular momentum in the parent particle rest frame, 
which decays into a two-particle helicity state $\ket{p\phi\theta\lambda}$. 
Here, $p=p_1-p_2$ and  $\lambda=\lambda_1-\lambda_2$ stand for the relative momenta and the helicity difference in the initial particle rest frame. 
Subscripts 1 and 2 denote the label of the produced particles. 
Notations and detailed derivations of the content given in this section can be found in Refs. \cite{Chung:1971ri,devanathan2005angular,Tung:1985na,Martin:1970hmp,Choi:2019aig}.

\subsection{Representation of the SO(3) algebra}\label{appendix:so3algebra}
We have already presented the rotation matrix parameterized by a rotation angle and rotation axis in Eq.(\ref{eq:rotmatrix}). 
It can alternatively be parameterized by 3 Euler angles $\alpha$, $\beta$ and $\gamma$, where $R(\alpha,\beta,\gamma)=e^{-i\alpha J_{3}}e^{-i\beta J_{2}}e^{-i\gamma J_{3}}$. 
From the commutation relations given in Eq.(\ref{eq:rotgen}), we can determine the Casimir operator as $\bm{J}^{2}=J_{1}^{2}+J_{2}^{2}+J_{3}^{2}$. Following the procedure given in elementary quantum mechanic textbooks, 
we find the irreducible representation of the SO(3) algebra to be labeled by eigenvalues of $\bm{J}^{2}$ and $J_3$, their eigenvectors satisfying the following orthogonality and completeness relations:
\begin{align}\label{eq:eigen}
\bm{J}^{2}\ket{jm}&=j(j+1)\ket{jm},\;J_{3}\ket{jm}=m\ket{jm}(m=-j,\dots,j),\nonumber\\
U\left(R(\alpha,\beta,\gamma)\right)\ket{jm}&=\sum_{m^{\prime}=-j}^{j}D^{j}\left(R(\alpha,\beta,\gamma)\right)_{m^{\prime}m}\ket{jm^{\prime}},\\
\braket{jm}{j^{\prime}m^{\prime}}&=\delta_{jj^{\prime}}\delta_{mm^{\prime}},\;\sum_{jm}\ketbra{jm}{jm}=1.\nonumber
\end{align}

\subsection{The Wigner D-matrix}
The Wigner $D$-matrix $D^{j}\left(R(\alpha,\beta,\gamma)\right)_{m^{\prime} m}$ is a matrix element of the irreducible representation of $R(\alpha,\beta,\gamma)$ defined in Eq.(\ref{eq:eigen}). 
\begin{align}\label{eq:wignerdmatrix}
D^{j}\left(R(\alpha,\beta,\gamma)\right)_{m^{\prime}m}&=\mel{jm^{\prime}}{U\left(R(\alpha,\beta,\gamma)\right)}{jm}=\mel{jm^{\prime}}{e^{-i\alpha J_{3}}e^{-i\beta J_{2}}e^{-i\gamma J_{3}}}{jm}\nonumber\\
&\\
&=e^{-i(m^{\prime}\alpha+m\gamma)}\mel{jm^{\prime}}{e^{-i\beta J_{2}}}{jm}\equiv e^{-i(m^{\prime}\alpha+m\gamma)}d^{j}(\beta)_{m^{\prime}m},\nonumber
\end{align}
where $d^{j}(\beta)$ is the Wigner $d$-matrix. $D^{j}\left(R(\alpha,\beta,\gamma)\right)_{m^{\prime} m}$ satisfies the following orthogonality and completeness relations,
\begin{align}\label{eq:orthocomple}
\int_{0}^{2\pi}\dd[]\alpha\int_{0}^{\pi}\dd[]\cos\beta\int_{0}^{2\pi}\dd[]\gamma D^{j\dagger}(\alpha,\beta,\gamma)_{mn}D^{j^{\prime}}(\alpha,\beta,\gamma)_{m^{\prime}n^{\prime}}&=\frac{8\pi^{2}}{2j+1}\delta_{jj^{\prime}}\delta_{mn^{\prime}}\delta_{nm^{\prime}},\nonumber\\
&\\
\sum_{jik}(2j+1)D^{j}(R)_{ik}D^{j\dagger}(R^{\prime})_{ki}&=\delta(R-R^{\prime}).\nonumber
\end{align}
More proofs regarding these orthogonality and completeness relations can be found in Ref.\cite{Tung:1985na}.

\subsection{One particle state of a massive particle}
We can describe a single particle state by using both the helicity and canonical bases. 
Initially, the particle is at rest so that its momentum $\bm{p}=\bm{0}$. The helicity state is constructed by a boost along $z$-direction by $\beta$ followed by an Euler rotation $R(\phi,\theta,0)$, so that the $z$-axis aligns with the particle momentum, specified as $\bm{p}=(p,\phi,\theta)$. The explicit transformation matrix is given as 
\begin{align}\label{eq:helboost}
h(\bm{\beta})&=R(\phi,\theta,0)L_3(\beta)=\begin{pmatrix}\gamma&0&0&\gamma\beta\\\gamma\beta\sin\theta\cos\phi&\cos\theta\cos\phi&-\sin\phi&\gamma\sin\theta\cos\phi\\\gamma\beta\sin\theta\sin\phi&\cos\theta\sin\phi&\cos\phi&\gamma\sin\theta\sin\phi\\\gamma\beta\cos\theta&-\sin\theta&0&\gamma\cos\theta\end{pmatrix}.
\end{align}
Here $\beta$ denotes the (conventionally normalized) particle velocity and $\gamma=1/\sqrt{1-\beta^2}$. 
The four-momentum of a particle in its rest frame transforms as
\begin{align}
p^\mu&=\left(M,\;0,\;0,\;0\right)\to\gamma M\left(1,\;\beta\sin\theta\cos\phi,\;\beta\sin\theta\sin\phi,\;\beta\cos\theta\right),
\end{align}
where $M$ is the particle mass. We label the state of the particle at rest by $J_3$ and define the general helicity 
state as
\begin{align}
\bm{J}^{2}\ket{\bm{0}j\lambda}&=j(j+1),\;J_{3}\ket{\bm{0}j\lambda}=\lambda\ket{\bm{0}j\lambda},\nonumber\\
&\\
\ket{\bm{p}j\lambda}&\equiv U\left(h(\bm{\beta})\right)\ket{\bm{0}j\lambda}=U\left(R(\phi,\theta,0)\right)\ket{p\hat{z}j\lambda},\nonumber
\end{align}
where $p=\gamma M\beta$. Since $\lambda$ is an eigenvalue of $J_3$, it is not altered by a Lorentz boost along the $z$-direction. Making use of the rotation invariance of helicity, it can be understood that $\lambda$ is also an eigenvalue of helicity operator, 
\begin{align}
\frac{\bm{J}\vdot\bm{P}}{\abs{\bm{P}}}\ket{\bm{p}j\lambda}&=U(R)U^{-1}(R)\frac{\bm{J}\vdot\bm{P}}{\abs{\bm{P}}}U(R)\ket{p\hat{z}j\lambda}=U(R)\frac{\bm{J}\vdot\bm{P}}{\abs{\bm{P}}}\ket{p\hat{z}j\lambda}=\lambda\ket{\bm{p}j\lambda}.
\end{align}
Since $\lambda$ is an eigenvalue of $J_3$, $\lambda=-j\dots j$. 

We can alternatively represent the state of a particle using the canonical basis, which is labeled by its momentum, total angular momentum and its $z$-component defined in the rest frame of a particle. 
A general canonical state with arbitrary momentum in the direction defined by the angles $(\phi,\theta)$ is constructed by first applying an inverse rotation, then performing a Lorentz boost along the $z$-axis, and finally rotating the particle back to its original configuration,
\begin{align}
\ket{\bm{p}jm}&\equiv U\left(L(\bm{\beta})\right)\ket{\bm{0}jm}=U\big(R(\phi,\theta,0)L_{3}(\beta)R^{-1}(\phi,\theta,0)\big)\ket{\bm{0}jm},\nonumber\\
&\\
L(\bm{\beta})&=\begin{pmatrix}\gamma&\gamma\beta_1&\gamma\beta_2&\gamma\beta_3\\\gamma\beta_1&1+\frac{\gamma-1}{\beta^{2}}\beta_{1}^{2}&\frac{\gamma-1}{\beta^{2}}\beta_{1}\beta_{2}&\frac{\gamma-1}{\beta^{2}}\beta_{1}\beta_{3}\\\gamma\beta_{2}&\frac{\gamma-1}{\beta^{2}}\beta_{1}\beta_{2}&1+\frac{\gamma-1}{\beta^{2}}\beta_{2}^{2}&\frac{\gamma-1}{\beta^{2}}\beta_{2}\beta_{3}\\\gamma\beta_{3}&\frac{\gamma-1}{\beta^{2}}\beta_{1}\beta_{3}&\frac{\gamma-1}{\beta^{2}}\beta_{2}\beta_{3}&1+\frac{\gamma-1}{\beta^{2}}\beta_{3}^{2}\end{pmatrix}.\nonumber
\end{align}
Here, $\bm{\beta}=\left(\beta_1,\;\beta_2,\;\beta_3\right)$ and $L(\bm{\beta})$ stands for a pure boost along the direction of $\bm{\beta}$. 
When the particle is at rest, the canonical state transforms under rotation as in Eq.(\ref{eq:eigen}). The connection between helicity and canonical state bases can be expressed as
\begin{align}\label{eq:helcanconnect}
\ket{\bm{p}j\lambda}&=U\big(R(\phi,\theta,0)L_{3}(\beta)R^{-1}(\phi,\theta,0)\big)U\big(R(\phi,\theta,0)\big)\ket{\bm{0}j\lambda}=\sum_{m=-j}^{j}D^{j}\left(R(\phi,\theta,0)\right)_{m\lambda}\ket{\bm{p}jm}.
\end{align}
We choose our normalization to be Lorentz invariant, such that 
\begin{align}
\braket{\bm{p}j\lambda}{\bm{p}^{\prime}j^{\prime}\lambda^{\prime}}&=(2\pi)^{3}2E_{\bm{p}}\delta^{3}(\bm{p}-\bm{p}^{\prime})\delta_{jj^{\prime}}\delta_{\lambda\lambda^{\prime}},\quad\braket{\bm{p}jm}{\bm{p}^{\prime}j^{\prime}m^{\prime}}=(2\pi)^{3}2E_{\bm{p}}\delta^{3}(\bm{p}-\bm{p}^{\prime})\delta_{jj^{\prime}}\delta_{mm^{\prime}}.
\end{align}
The transformation of the Dirac delta function under a Lorentz boost can be derived by making use of the property $\delta\left(f(x)\right)=\sum_{i}\delta(x-x_{i})/\abs{f^{\prime}(x_{i})}$. 
Let $\bm{p}$ and $\bm{q}$ be related by a Lorentz transformation such that $E_{p}=\gamma(E_{q}+\beta q),\;p=\gamma(q+\beta E_q)$. 
We can then prove that the delta function $E_{p}\delta(p)$ is Lorentz invariant, as
\begin{align}
\delta(p)&=\frac{1}{\abs{\dv{p}{q}}}\delta(q)=\frac{1}{\gamma(1+\beta\frac{q}{E_q})}\delta(q)=\frac{E_q}{E_p}\delta(q).
\end{align}

\subsection{Two particle state}\label{appendix:twoparticle}
By definition, a two particle helicity state is defined as a tensor product of two one particle states in the c.m. frame,  
\begin{align}
\ket{\phi\theta\lambda_{1}\lambda_{2}}&=U\big(R(\phi,\theta,0)\big)\big[U\big(L_{3}(\beta)\big)\ket{\bm{0}j_{1}\lambda_{1}}\otimes U\big(L_{3}(-\beta)\big)\ket{\bm{0}j_{2}-\lambda_{2}}\big].
\end{align}
To calculate its angular decay distribution, we need to determine the relationship between a two particle helicity state and $\ket{JM\lambda_{1}\lambda_{2}}$. 
Here, $J$ and $M$ stand for the total angular momentum and its $z$-th component of the initial particle. 
We assume these two different states are connected by a coefficient $C_{JM}(\phi,\theta,\lambda_{1},\lambda_{2})$, 
\begin{align}
\ket{\phi\theta\lambda_{1}\lambda_{2}}&=\sum_{JM}C_{JM}\big(\phi,\theta,\lambda_{1},\lambda_{2}\big)\ket{JM\lambda_{1}\lambda_{2}}.
\label{eq:derive_C_1}
\end{align}
To evaluate the coefficient $C_{JM}(\phi,\theta,\lambda_{1},\lambda_{2})$, we first make use of the standard helicity state in which $\phi=\theta=0$, 
\begin{align}
\ket{00\lambda_{1}\lambda_{2}}&=\sum_{JM}C_{JM}(0,0,\lambda_{1},\lambda_{2})\ket{JM\lambda_{1}\lambda_{2}}=\sum_{J}C_{J\lambda}(0,0,\lambda_{1},\lambda_{2})\ket{J\lambda\lambda_{1}\lambda_{2}}.
\end{align}
When $\phi=\theta=0$, particle 1 is emitted in the forward $(+z)$ direction while particle 2 is emitted backwards $(-z)$. 
Their projection of angular momentum onto the $z$-axis then becomes $\lambda=\lambda_{1}-\lambda_{2}$. 
We can also view the two particle helicity state as it is rotated from a standard state. Thus, 
\begin{align}
\ket{\phi\theta\lambda_{1}\lambda_{2}}&=U(R)\ket{00\lambda_{1}\lambda_{2}}=\sum_{J}C_{J\lambda}(0,0,\lambda_{1},\lambda_{2})U(R)\ket{J\lambda\lambda_{1}\lambda_{2}}=\sum_{JM}C_{J\lambda}(0,0,\lambda_{1},\lambda_{2})D^{J}(R)_{M\lambda}\ket{JM\lambda_{1}\lambda_{2}}.
\label{eq:derive_C_2}
\end{align}
In the last line, we have used the fact that a state of definite angular momentum transforms as a canonical state does in Eq.(\ref{eq:helcanconnect}). 
We furthermore take advantage of the orthogonality and completeness relation in Eq.(\ref{eq:orthocomple}) of the Wigner $D$-matrix and of the canonical state. 
Therefore, we can obtain $C_{JM}(0,0,\lambda_{1},\lambda_{2})$ as 
\begin{align}
C_{J\lambda}(0,0,\lambda_{1},\lambda_{2}) = \sqrt{\frac{2J+1}{4\pi}}.
\end{align}
By comparing Eq.(\ref{eq:derive_C_1}) with the last line of Eq.(\ref{eq:derive_C_2}), we obtain the final results as
\begin{align}\label{eq:derive_C_3}
C_{JM}(\phi,\theta,\lambda_{1},\lambda_{2}) = \sqrt{\frac{2J+1}{4\pi}} D^{J}(R)_{M\lambda}.
\end{align}
Here, $\lambda = \lambda_{1}-\lambda_{2}$.

\subsection{Two body decay amplitude}\label{appendix:decayamplitude}
When a particle decays, the initial state of definite angular momentum $\ket{JM}$ transits to a final two particle helicity state $\ket{\phi\theta\lambda_{1}\lambda_{2}}$. The two body decay amplitude is then evaluated as 
\begin{align}
f_{\lambda M}&=\mel{\phi\theta\lambda_{1}\lambda_{2}}{H_{\text{int}}}{JM}=\sum_{J'M'\lambda'_{1}\lambda'_{2}}\braket{\phi\theta\lambda_{1}\lambda_{2}}{J'M'\lambda_{1}'\lambda_{2}'}\mel{J'M'\lambda_{1}'\lambda_{2}'}{H_{\text{int}}}{JM}\nonumber\\
&=\sqrt{\frac{2J+1}{4\pi}}D^{J}(R)^{*}_{M\lambda}\mel{JM\lambda_{1}\lambda_{2}}{H_{\text{int}}}{JM}, 
\label{eq:transition_amplitude}
\end{align}
where in the second line, we have used Eq.(\ref{eq:derive_C_3}) and angular momentum conservation. 
$H_{\text{int}}$ denotes the interaction Hamiltonian of the decay process. 
The matrix element $\mel{JM\lambda_{1}\lambda_{2}}{H_{\text{int}}}{JM}$ depends on the rotation invariant quantities $J$, $\lambda_{1}$ and $\lambda_{2}$, but is independent of $M$. 
Thus, we denote it as $\mel{JM\lambda_{1}\lambda_{2}}{H_{\text{int}}}{JM}\equiv H^{J}_{\text{int}}(\lambda_{1},\lambda_{2})$. Let us specify the initial state as $\ket{I}=\sum_{M}a_{M}\ket{JM}$ and define the spin density matrix as in Eq.(\ref{eq:vec_meson_config}), $\rho_{MM^{\prime}} = a_{M} a^{*}_{M^{\prime}}$. 
We can then calculate the normalized angular distribution as
$I(\phi,\theta) = \abs{\sum_{M} a_{M} f_{\lambda M}}^2/\Gamma$, in which $\Gamma$ is the decay width 
of the initial particle. Finally, we can write down the general formula as
\begin{align}
I(\phi,\theta)&=\frac{1}{\Gamma}\sum_{\lambda_{1}\lambda_{2}}\sum_{MM^{\prime}}f_{\lambda M}\rho_{MM^{\prime}}f^{*}_{\lambda M^{\prime}}=\frac{1}{\Gamma}\sum_{\lambda_{1}\lambda_{2}}\sum_{MM^{\prime}}\mel{\phi\theta\lambda_{1}\lambda_{2}}{H_{\text{int}}}{JM}\rho_{MM^{\prime}}\mel{JM^{\prime}}{H_{\text{int}}}{\phi\theta\lambda_{1}\lambda_{2}}\\
&=\frac{1}{\Gamma}\sum_{\lambda_{1}\lambda_{2}}\sum_{MM^{\prime}}\frac{2J+1}{4\pi}D^{J\dagger}\tensor{(\phi,\theta,0)}{_\lambda_M}\rho_{MM^{\prime}}D^{J}\tensor{(\phi,\theta,0)}{_{M^{\prime}}_\lambda}\abs{H^{J}_{\text{int}}(\lambda_{1},\lambda_{2})}^{2}.\nonumber
\end{align}

For the $V\to PP$ decay, the only matrix element we need is $H^{1}_{\text{int}}(0,0)$. Hence, the angular distribution is solely determined by the rotation matrix $D^{1}(\phi,\theta,0)$ as 
\begin{align}
I(\phi,\theta)&=\frac{3}{8\pi}\Bigg(1-\rho_{00}+(3\rho_{00}-1)\cos^{2}\theta-2\text{Re}[\rho_{1-1}]\sin^{2}\theta\cos2\phi+2\text{Im}[\rho_{1-1}]\sin^{2}\theta\sin2\phi\\
&-\sqrt{2}\text{Re}[\rho_{10}-\rho_{-10}]\sin2\theta\cos\phi+\sqrt{2}\text{Im}[\rho_{10}+\rho_{-10}]\sin2\theta\sin\phi\Bigg).\nonumber
\end{align}
This fits the earlier result given in Eq.(\ref{eq:vppangledist}). 

For the $V\to ll$ decay, the matrix element $H_{\text{int}}^{1}(1/2,-1/2)=H_{\text{int}}^{1}(-1/2,1/2)$ is needed to evaluate the angular distribution. 
The above equality sign holds thanks to parity conservation in the electromagnetic decay \cite{Chung:1971ri}. 
The corresponding distribution is obtained as 
\begin{align}
I(\phi,\theta)&=\frac{3}{16\pi}\bigg(1+\rho_{00}+(1-3\rho_{00})\cos^{2}\theta+2\text{Re}[\rho_{1-1}]\sin^{2}\theta\cos2\phi-2\text{Im}[\rho_{1-1}]\sin^{2}\theta\sin2\phi\\
&+\sqrt{2}\text{Re}[\rho_{10}-\rho_{-10}]\sin2\theta\cos\phi-\sqrt{2}\text{Im}[\rho_{10}+\rho_{-10}]\sin2\theta\sin\phi\bigg),\nonumber
\end{align}
where this agrees with Eq.(\ref{eq:vllangledist}).

Finally, for the $K_{1}\to K^{*}\pi$ decay, we need to consider the 3 matrix elements $H^{1}_{\text{int}}(1,0)$, $H^{1}_{\text{int}}(-1,0)$ 
and $H^{1}_{\text{int}}(0,0)$. Since the decay is parity conserving, $H^{1}_{\text{int}}(1,0)=H^{1}_{\text{int}}(-1,0)$. The interaction Hamiltonian can easily be obtained from the interaction Lagrangian given in Eq.(\ref{eq:k1lagrangian}) as $H_{\text{int}}=-L_{\text{int}}$. The two body decay amplitude of Eq.(\ref{eq:transition_amplitude}) can be calculated using the polarization vectors given in Appendix \ref{appendix:k1angdist}. The relative strength of the two terms is determined by evaluating 
two independent transition amplitudes, with an outgoing vector particle carrying a different helicity $\lambda_{1}$. 
Specifically, we have
\begin{align}
f_{1 1} \propto \epsilon^{\ast}_{K^*}(1) \cdot \epsilon_{K_1}(1) = -\frac{1 + \cos \theta}{2} e^{i\phi}, 
\label{eq:transition_amplitude_11}
\end{align}
and 
\begin{align}
f_{0 1} \propto \epsilon^{\ast}_{K^*}(0) \cdot \epsilon_{K_1}(1) = -\frac{E_{1}}{\sqrt{2}m_{1}} \sin\theta e^{i\phi}.  
\label{eq:transition_amplitude_01}
\end{align}
Comparing this with Eq. (\ref{eq:transition_amplitude}), we find that 
\begin{align}
H^{1}_{\text{int}}(1,0) = H^{1}_{\text{int}}(-1,0) \propto g_{K_{1}K^*\pi},\quad H^{1}_{\text{int}}(0,0) \propto g_{K_{1}K^*\pi}\frac{E_{1}}{m_{1}}.
\end{align}
Now, we have sufficient information to calculate the angular distribution of the $K_{1}\to K^{*}\pi$ decay as
\begin{align}
I(\phi,\theta)&=\frac{1}{\Gamma}\frac{3}{4\pi}\Bigg(\abs{H(1,0)}^{2}+\frac{1}{2}\Big(\abs{H(0,0)}^{2}-\abs{H(1,0)}^{2}\Big)\nonumber\\
&\Big(1-\rho_{00}+(3\rho_{00}-1)\cos^{2}\theta-2\text{Re}[\rho_{1-1}]\sin^{2}\theta\cos2\phi+2\text{Im}[\rho_{1-1}]\sin^{2}\theta\sin2\phi\nonumber\\
&-\sqrt{2}\text{Re}[\rho_{10}-\rho_{-10}]\sin2\theta\cos\phi+\sqrt{2}\text{Im}[\rho_{10}+\rho_{-10}]\sin2\theta\sin\phi\Big)\Bigg)\\
&=\frac{3}{4\pi\left(3+\frac{\abs{\bm{p_{1}}}^{2}}{m_{1}^{2}}\right)}\Bigg(1+\frac{\abs{\bm{p_{1}}}^{2}}{2m_{1}^2}\Big(1-\rho_{00}+(3\rho_{00}-1)\cos^{2}\theta-2\text{Re}[\rho_{1-1}]\sin^{2}\theta\cos2\phi+2\text{Im}[\rho_{1-1}]\sin^{2}\theta\sin2\phi\nonumber\\
&-\sqrt{2}\text{Re}[\rho_{10}-\rho_{-10}]\sin2\theta\cos\phi+\sqrt{2}\text{Im}[\rho_{10}+\rho_{-10}]\sin2\theta\sin\phi\Big)\Bigg), \nonumber
\end{align}
which agrees with Eq.(\ref{eq:avpangledist}).


\begin{thebibliography}{99}

\bibitem{Nambu:1961tp}
Y.~Nambu and G.~Jona-Lasinio,
%``Dynamical Model of Elementary Particles Based on an Analogy with Superconductivity. 1.,''
Phys. Rev. \textbf{122}, 345 (1961).
%doi:10.1103/PhysRev.122.345
%5451 citations counted in INSPIRE as of 14 Oct 2020

%\cite{Nambu:1961fr} 
\bibitem{Nambu:1961fr}
Y.~Nambu and G.~Jona-Lasinio,
%``DYNAMICAL MODEL OF ELEMENTARY PARTICLES BASED ON AN ANALOGY WITH SUPERCONDUCTIVITY. II,''
Phys. Rev. \textbf{124}, 246 (1961).
%doi:10.1103/PhysRev.124.246
%2740 citations counted in INSPIRE as of 14 Oct 2020

%\cite{Hatsuda:1985eb} 
\bibitem{Hatsuda:1985eb}
  T.~Hatsuda and T.~Kunihiro,
  %``Fluctuation Effects in Hot Quark Matter: Precursors of Chiral Transition at Finite Temperature,''
  Phys.\ Rev.\ Lett.\  {\bf 55}, 158 (1985).
  %doi:10.1103/PhysRevLett.55.158
  %%CITATION = doi:10.1103/PhysRevLett.55.158;%%
  %329 citations counted in INSPIRE as of 05 Aug 2016

%\cite{Brown:1991kk} 
\bibitem{Brown:1991kk}
  G.~E.~Brown and M.~Rho,
  %``Scaling effective Lagrangians in a dense medium,''
  Phys.\ Rev.\ Lett.\  {\bf 66}, 2720 (1991).
  %doi:10.1103/PhysRevLett.66.2720
  %%CITATION = doi:10.1103/PhysRevLett.66.2720;%%
  %1223 citations counted in INSPIRE as of 05 Aug 2016

%\cite{Hatsuda:1991ez} 
\bibitem{Hatsuda:1991ez}
  T.~Hatsuda and S.~H.~Lee,
  %``QCD sum rules for vector-mesons in the nuclear medium,''
  Phys.\ Rev.\ C {\bf 46}, R34 (1992).
  %doi:10.1103/PhysRevC.46.R34
  %%CITATION = doi:10.1103/PhysRevC.46.R34;%%
  %671 citations counted in INSPIRE as of 05 Aug 2016


%\cite{Leupold:2009kz} 
\bibitem{Leupold:2009kz}
  S.~Leupold, V.~Metag, and U.~Mosel,
  %``Hadrons in strongly interacting matter,''
  Int.\ J.\ Mod.\ Phys.\ E {\bf 19}, 147 (2010).
  %doi:10.1142/S0218301310014728
%  [arXiv:0907.2388 [nucl-th]].
  %%CITATION = doi:10.1142/S0218301310014728;%%
  %139 citations counted in INSPIRE as of 04 Aug 2016

%\cite{Gubler:2014pta}
\bibitem{Gubler:2014pta}
P.~Gubler and K.~Ohtani,
%``Constraining the strangeness content of the nucleon by measuring the \ensuremath{\phi} meson mass shift in nuclear matter,''
Phys. Rev. D \textbf{90}, no.9, 094002 (2014).
%doi:10.1103/PhysRevD.90.094002
%[arXiv:1404.7701 [hep-ph]].
%50 citations counted in INSPIRE as of 26 Sep 2024

\bibitem{Hayano:2008vn}
For review see, R.~S.~Hayano and T.~Hatsuda,
  %``Hadron properties in the nuclear medium,''
  Rev.\ Mod.\ Phys.\  {\bf 82}, 2949 (2010).
 % doi:10.1103/RevModPhys.82.2949
%  [arXiv:0812.1702 [nucl-ex]].

%\cite{JPARC:2023quf}
\bibitem{JPARC:2023quf}
M.~Ichikawa \textit{et al.,}
%``Commissioning Runs of J-PARC E16 Experiment,''
Acta Phys. Pol. B Proc. Suppl. \textbf{16}, 1-A143 (2023). 
%https://doi.org/10.5506/APhysPolBSupp.16.1-A143.
%2 citations counted in INSPIRE as of 24 Mar 2024


%\cite{Metag:2017yuh} 
\bibitem{Metag:2017yuh}
  V.~Metag, M.~Nanova, and E.~Y.~Paryev,
  %``Meson\UTF{2013}nucleus potentials and the search for meson\UTF{2013}nucleus bound states,''
  Prog.\ Part.\ Nucl.\ Phys.\  {\bf 97}, 199 (2017).
%  doi:10.1016/j.ppnp.2017.08.002
%  [arXiv:1706.09654 [nucl-ex]].
  %%CITATION = doi:10.1016/j.ppnp.2017.08.002;%%
  %14 citations counted in INSPIRE as of 01 Nov 2018


%\cite{Ohnishi:2019cif} 
\bibitem{Ohnishi:2019cif}
H.~Ohnishi, F.~Sakuma, and T.~Takahashi,
%``Hadron Physics at J-PARC,''
Prog. Part. Nucl. Phys. \textbf{113}, 103773 (2020).
%doi:10.1016/j.ppnp.2020.103773
%[arXiv:1912.02380 [nucl-ex]].
%2 citations counted in INSPIRE as of 18 Oct 2020

%\cite{Salabura:2020tou} 
\bibitem{Salabura:2020tou}
P.~Salabura and J.~Stroth,
%``Dilepton radiation from strongly interacting systems,''
Prog. Part. Nucl. Phys. \textbf{120}, 103869 (2021).
%28 citations counted in INSPIRE as of 19 Jun 2024



%\cite{LeBellac:1996}
\bibitem{LeBellac:1996}
M.~Le Bellac,
\textit{Thermal Field Theory}
(Cambridge University Press, Cambridge, England, 1996).



\bibitem{Lee:1997zta}
S.~H.~Lee,
%``vector-mesons in-medium with finite three momentum,''
Phys. Rev. C \textbf{57}, 927  (1998)
; \textbf{58}, 3771(E)(1998).
%doi:10.1103/PhysRevC.57.927
%48 citations counted in INSPIRE as of 27 Feb 2024

%\cite{Kim:2019ybi}
\bibitem{Kim:2019ybi}
H.~Kim and P.~Gubler,
%``The \ensuremath{\phi} meson with finite momentum in a dense medium,''
Phys. Lett. B \textbf{805}, 135412 (2020).
%doi:10.1016/j.physletb.2020.135412
%19 citations counted in INSPIRE as of 27 Feb 2024



%\cite{Lee:2023ofg}
\bibitem{Lee:2023ofg}
S.~H.~Lee,
%``Chiral Symmetry Breaking and the Masses of Hadrons: A Review,''
Symmetry \textbf{15}, 799 (2023).
%doi:10.3390/sym15040799
%2 citations counted in INSPIRE as of 27 Feb 2024

\bibitem{Aoki:2023qgl}
K.~Aoki \textit{et}. al.,
%Experimental Study of In-medium Spectral Change of vector-mesons at J-PARC,
%doi:10.1007/s00601-023-01828-7
Few-Body Syst. \textbf{64}, 63 (2023).

%\cite{KEK-PS-E325:2005wbm}
\bibitem{KEK-PS-E325:2005wbm}
R.~Muto \textit{et al.} [KEK-PS-E325],
%``Evidence for in-medium modification of the phi meson at normal nuclear density,''
Phys. Rev. Lett. \textbf{98}, 042501 (2007).
%doi:10.1103/PhysRevLett.98.042501
%[arXiv:nucl-ex/0511019 [nucl-ex]].
%190 citations counted in INSPIRE as of 24 Sep 2024



\bibitem{Sako}
H.~Sako \textit{et} al.,
"P88: Study of in-medium modification of $\phi$ mesons inside the nucleus with $\phi \rightarrow K^+ K^-$ measurement with the E16 spectrometer",
https://j-parc.jp/researcher/Hadron/en/pac\_2107/pdf/P88\_2021-12.pdf.

%\cite{Lee:2019tvt}
\bibitem{Lee:2019tvt}
S.~H.~Lee,
%``Theory on Hadrons in Nuclear Medium,''
JPS Conf. Proc. \textbf{26}, 011012 (2019).
%doi:10.7566/JPSCP.26.011012
%9 citations counted in INSPIRE as of 27 Feb 2024

%\cite{Song:2018plu}
\bibitem{Song:2018plu}
T.~Song, T.~Hatsuda, and S.~H.~Lee,
%``QCD sum rule for open strange meson $K_1^\pm$ in nuclear matter,''
Phys. Lett. B \textbf{792}, 160 (2019).
%doi:10.1016/j.physletb.2019.03.023
%16 citations counted in INSPIRE as of 01 Mar 2024

%\cite{STAR:2017ckg}
\bibitem{STAR:2017ckg}
L.~Adamczyk \textit{et al.}
``(STAR Collaboration), Global $\Lambda$ hyperon polarization in nuclear collisions: Evidence for the most vortical fluid,''
Nature (London) \textbf{548}, 62 (2017).
%doi:10.1038/nature23004
%496 citations counted in INSPIRE as of 03 Aug 2022

%\cite{STAR:2018gyt}
\bibitem{STAR:2018gyt}
J.~Adam \textit{et al.}
``(STAR Collaboration), Global polarization of $\Lambda$ hyperons in Au+Au collisions at $\sqrt{s_{_{NN}}}$ = 200 GeV,''
Phys. Rev. C \textbf{98}, 014910 (2018).
%doi:10.1103/PhysRevC.98.014910
%215 citations counted in INSPIRE as of 03 Aug 2022

%\cite{STAR:2021beb}
\bibitem{STAR:2021beb}
M.~S.~Abdallah \textit{et al.} ''(STAR Collaboration),
Global $\Lambda$-hyperon polarization in Au+Au collisions at $\sqrt {s_{NN}}$=3~GeV,''
Phys. Rev. C \textbf{104}, L061901 (2021).
%doi:10.1103/PhysRevC.104.L061901
%29 citations counted in INSPIRE as of 05 Aug 2022

%\cite{HADES:2021Kornas}
\bibitem{HADES:2021Kornas}
F.~J.~Kornas,
``Global polarization of {\ensuremath{\Lambda}} hyperons as a probe for vortical effects in A+A collisions at HADES,''
%doi:10.26083/tuprints-00019763
Ph.D. thesis, Technische University Darmstadt, 2021.



%\cite{STAR:2020xbm}
\bibitem{STAR:2020xbm}
J.~Adam \textit{et al.}, 
%``(STAR Collaboration), Global Polarization of $\Xi$ and $\Omega$ Hyperons in Au+Au Collisions at $\sqrt {s_{NN}}$ = 200  GeV,''
Phys. Rev. Lett. \textbf{126}, 162301 (2021).
%doi:10.1103/PhysRevLett.126.162301
%32 citations counted in INSPIRE as of 05 Aug 2022

%\cite{Park:2022ayr}
\bibitem{Park:2022ayr}
I.~W.~Park, H.~Sako, K.~Aoki, P.~Gubler and S.~H.~Lee,
%``Disentangling longitudinal and transverse modes of the \ensuremath{\phi} meson through dilepton and kaon decays,''
Phys. Rev. D \textbf{107}, no.7, 074033 (2023).
%doi:10.1103/PhysRevD.107.074033
%[arXiv:2211.16949 [hep-ph]].
%11 citations counted in INSPIRE as of 26 Sep 2024

%\cite{Park:2024vga}
\bibitem{Park:2024vga}
I.~W.~Park, H.~Sako, K.~Aoki, P.~Gubler and S.~H.~Lee,
%``Identifying the transverse and longitudinal modes of the K* and K1 mesons through their angular-dependent decay modes,''
Phys. Rev. D \textbf{109}, no.11, 114042 (2024).
%doi:10.1103/PhysRevD.109.114042
%[arXiv:2403.18288 [hep-ph]].
%1 citations counted in INSPIRE as of 26 Sep 2024

%\cite{Faccioli:2010kd}
\bibitem{Faccioli:2010kd}
P.~Faccioli, C.~Lourenco, J.~Seixas and H.~K.~Wohri,
%``Towards the experimental clarification of quarkonium polarization,''
Eur. Phys. J. C \textbf{69}, 657 (2010).
%doi:10.1140/epjc/s10052-010-1420-5
%166 citations counted in INSPIRE as of 14 Jul 2022

 %%--------------------%--------------------------------------------------------------

%\cite{Schilling:1969um}
\bibitem{Schilling:1969um}
K.~Schilling, P.~Seyboth, and G.~E.~Wolf,
%``On the Analysis of vector-meson Production by Polarized Photons,''
Nucl. Phys. \textbf{B 15}, 397 (1970)
\textbf{B18}, 332(E) (1970).
%doi:10.1016/0550-3213(70)90070-2
%340 citations counted in INSPIRE as of 26 Feb 2024

%\cite{Workman:2022ynf}
\bibitem{Workman:2022ynf}
R.~L.~Workman \textit{et al.} (Particle Data Group),
%``Review of Particle Physics,''
Prog. Theor .Exp. Phys.\textbf{2022}, 083C01 (2022).
%doi:10.1093/ptep/ptac097
%3 citations counted in INSPIRE as of 12 Jul 2022

%\cite{Klingl:1996by}
\bibitem{Klingl:1996by}
F.~Klingl, N.~Kaiser, and W.~Weise,
%``Effective Lagrangian approach to vector-mesons, their structure and decays,''
Z. Phys. A \textbf{356}, 193 (1996). 
%doi:10.1007/s002180050167
%233 citations counted in INSPIRE as of 31 Dec 2021



%\cite{Kaymakcalan:1984bz}
\bibitem{Kaymakcalan:1984bz}
O.~Kaymakcalan and J.~Schechter, 
%``Chiral Lagrangian of pseudoscalars and vectors,'' 
Phys. Rev. D \textbf{31}, 1109 (1985).
%doi:10.1103/PhysRevD.31.1109
%207 citations counted in INSPIRE as of 27 Feb 2023

%\cite{Meissner:1987ge}
\bibitem{Meissner:1987ge}
U.~G.~Meissner, 
%Low-energy hadron physics from effective chiral Lagrangians with vector-mesons,
Phys. Rep. \textbf{161}, 213 (1988). 
%doi:10.1016/0370-1573(88)90090-7
%792 citations counted in INSPIRE as of 27 Feb 2023



%\cite{Sung:2021myr}
\bibitem{Sung:2021myr}
H.~S.~Sung, S.~Cho, J.~Hong, S.~H.~Lee, S.~Lim, and T.~Song,
%``K1/K\textasteriskcentered{} enhancement as a signature of chiral symmetry restoration in heavy ion collisions,''
Phys. Lett. B \textbf{819}, 136388 (2021).
%doi:10.1016/j.physletb.2021.136388
%2 citations counted in INSPIRE as of 25 Feb 2024



\bibitem{sakurai1969currents}
J.J. Sakurai, \textit{Currents and Mesons}, (University of Chicago Press, Chicago 1969).



\bibitem{Wigner1959}
E.P. Wigner, \textit{Group Theory and its Application to the Quantum Mechanics of Atomic Spectra} (Academic Press, Reading, MA, 1959).

%\cite{Leader:2011vwq}
\bibitem{Leader:2011vwq}
E. Leader, Cambridge Monogr. Part. Phys. Nucl. Phys. Cosmol. \textbf{15}, 1 (2023).
%13 citations counted in INSPIRE as of 08 May 2024



%\cite{Chung:1971ri}
\bibitem{Chung:1971ri}
S.~U.~Chung, Spin formalisms, Report No. CERN-71-08. %https://doi.org/10.5170/CERN-1971-008 (1971).
%78 citations counted in INSPIRE as of 27 Feb 2024

\bibitem{devanathan2005angular}
V.~Devanathan,
\textit{Angular Momentum Techniques in Quantum Mechanics}, 
%https://doi.org/10.1007/0-306-47123-x, 
(Springer, Netherlands, 2005).

%\cite{Tung:1985na}
\bibitem{Tung:1985na}
W.~K.~Tung,
``GROUP THEORY IN PHYSICS,''
%9 citations counted in INSPIRE as of 09 Aug 2024

%\cite{Martin:1970hmp}
\bibitem{Martin:1970hmp}
A.~D.~Martin and T.~D.~Spearman,
``Elementary Particle Theory,''
North-Holland Publishing Co., 1970,
ISBN 978-0-7204-0157-8
%9 citations counted in INSPIRE as of 09 Aug 2024

%\cite{Choi:2019aig}
\bibitem{Choi:2019aig}
S.~Y.~Choi, J.~H.~Jeong and J.~H.~Song,
%``General Spin Analysis from Angular Correlations in Two-Body Decays,''
Eur. Phys. J. Plus \textbf{135}, no.2, 210 (2020).
%doi:10.1140/epjp/s13360-020-00132-1
%[arXiv:1903.00166 [hep-ph]].
%4 citations counted in INSPIRE as of 06 Sep 2024


\end{thebibliography}
\end{document}